\documentclass{IEEEtran}
\usepackage{url}
\usepackage{amsmath}
\usepackage{amssymb}
\usepackage{amsthm}
\usepackage{authblk}
\usepackage{upgreek}
\usepackage[ruled,algonl,lined]{algorithm2e}
\usepackage{cite}
\usepackage[dvipsnames]{xcolor}
\usepackage{comment}
\usepackage{enumitem}

\newtheorem*{corollary*}{Corollary}

\usepackage[normalem]{ulem}

\DeclareMathAlphabet{\mathbit}{OML}{cmr}{bx}{it}
\DeclareMathAlphabet{\mathsf}{OT1}{cmss}{m}{n}
\DeclareMathAlphabet{\mathbsf}{OT1}{cmss}{bx}{it}

\newcommand{\inset}[2]{\ensuremath{\in \left\{#1,\ldots,#2\right\}}}

\newcommand*{\defeq}{\mathrel{\vcenter{\baselineskip0.5ex \lineskiplimit0pt
\hbox{\scriptsize.}\hbox{\scriptsize.}}}%
=}

\newcommand{\diff}{\ensuremath{\:\mathrm{d}}}

\DeclareMathOperator{\sinc}{sinc}

\newlength{\figurewidth}
\newlength{\figureheight}

\usepackage{tikz}
\usetikzlibrary{arrows,decorations,backgrounds,shadows,plotmarks,calc, positioning}
\usetikzlibrary{arrows.meta}

\usepackage{pgfplots}
\pgfplotsset{compat=newest}
\pgfplotsset{plot coordinates/math parser=false}
\pgfplotsset{every axis/.append style={font=\footnotesize}}
\pgfplotsset{
	ylabel right/.style={
		after end axis/.append code={
			\node [rotate=90, anchor=north] at (rel axis cs:1,0.5) {#1};
		}   
	}
}

\usepackage[hidelinks]{hyperref}

\usepackage{fancyhdr}
\fancypagestyle{others}{%
\chead{\scriptsize This is the author's version of an article that has been published in this journal. Changes were made to this version by the publisher prior to publication. \\
The final version of record is available at 	\textcolor{blue}{\url{http://dx.doi.org/10.1109/JLT.2021.3069686}}
}
\rhead{\thepage}
\cfoot{\scriptsize Copyright (c) 2021 IEEE. Personal use is permitted. For any other purposes, permission must be obtained from the IEEE by emailing pubs-permissions@ieee.org.
}
}
\title{Mismatched Models to Lower Bound the Capacity of Dual-Polarization Optical Fiber Channels}
\author{Francisco Javier Garc\'ia-G\'omez and Gerhard Kramer, \IEEEmembership{Fellow, IEEE}
\IEEEcompsocitemizethanks{
	\IEEEcompsocthanksitem
	Date of current version \today. 
	This work was supported by the German Research Foundation (DFG) under Grants KR 3517/8-1 and 3517/8-2.
	(\emph{Corresponding author: Francisco Javier Garc\'ia-G\'omez.})
	\\ \indent
	The authors are with the Institute for Communications Engineering, Technical University of Munich, 80333 Munich, Germany (e-mail: javier.garcia@tum.de; gerhard.kramer@tum.de).
	}
	}  

\begin{document}
\maketitle


\begin{abstract}
Regular perturbation is applied to the Manakov equation and motivates a generalized correlated phase-and-additive noise model for wavelength-division multiplexing over dual-polarization optical fiber channels. The model includes three hidden Gauss-Markov processes: phase noise, polarization rotation, and additive noise. Particle filtering is used to compute lower bounds on the capacity of multi-carrier communication with frequency-dependent powers and delays. A gain of 0.17 bits/s/Hz/pol in spectral efficiency or 0.8 dB in power efficiency is achieved with respect to existing models at their peak data rate. Frequency-dependent delays also increase the spectral efficiency of single-polarization channels.
\end{abstract}


\thispagestyle{others}
\pagestyle{others}

\section{Introduction}
Dual-polarization (2-pol) transmission almost doubles the data rates of optical fiber links~\cite{herard1991new, fludger2008coherent}. The apparent small reduction from a factor of two is caused by nonlinear coupling of polarizations due to the Kerr effect~\cite{agrawal_nfo,secondini2019nonlinearity}. A spectral efficiency upper bound of $\log_2(1+\textrm{SNR})$ bits/s/Hz/pol follows by generalizing~\cite{kramer2015upper, yousefi2015upper} to the 2-pol Manakov equation, where SNR is the receiver signal-to-noise ratio.

Simplified versions of the nonlinear Schr\"odinger equation (NLSE) can serve as mismatched models to compute capacity lower bounds for 1-pol channels, see the review in~\cite{ghozlan2017models}. For example, capacity bounds for 1-pol channels are derived in~\cite{essiambre_limits} by using wavelength-division multiplexing (WDM) and a conditionally Gaussian model where the noise variance depends on the amplitude of the transmitted symbol. The Gaussian Noise~\cite{poggiolini2014gn} and Enhanced Gaussian Noise~\cite{carena2014egn} models refine this approach, also for 2-pol channels, by giving analytical expressions for the power spectral density (PSD) of the nonlinear interference that is assumed to be Gaussian.

The regular perturbation (RP) model~\cite{mecozzi2000analysis, mecozzi2000system, mecozzi2001cancellation, vannucci2002rp} of the NLSE leads to correlated phase noise models~\cite{mecozzi2012nonlinear, dar2013properties, dar2014new} and improved capacity bounds for 1-pol. There is less literature on RP for 2-pol: we have found analyses only for special input signals such as Gaussian pulses~\cite{tao2011multiplier} or Fourier series with random Gaussian coefficients~\cite{johannisson2013perturbation}. A logarithmic perturbation (LP) model is developed in~\cite{secondini2012analytical, secondini2013achievable} that suggests a time- and frequency-varying polarization and phase noise (PPN) model. The latter model, together with WDM and multiple carriers per wavelength, was used in~\cite{secondini2019nonlinearity} to compute the best 2-pol capacity bounds that we are aware of.

We used RP in~\cite{garcia2020mismatched} to develop a correlated phase-and-additive noise (CPAN) model for WDM for 1-pol transmission. The model improved the capacity bounds in~\cite{secondini2017fiber} by applying a whitening filter and multi-carrier communication with frequency-dependent power allocation (FDPA). We here extend the CPAN model to 2-pol. The proposed 2pCPAN model includes phase noise, a random polarization rotation, and additive noise. All three impairments are correlated in time and the phase noise is correlated across polarizations. Using multiple carriers per wavelength, FDPA, and frequency-dependent delays, 2pCPAN improves the rates in~\cite{secondini2019nonlinearity} by 0.17 bits/s/Hz/pol in spectral efficiency or 0.8 dB in power efficiency at the peak data rate.

This paper is organized as follows. Sec.~\ref{sec:prelim} describes notation and the 2-pol propagation model. Sec.~\ref{sec:rp} develops continuous- and discrete-time RP models. Sec.~\ref{sec:2pcpan} presents the 2pCPAN model and Sec.~\ref{sec:simplified} simplifies the model to make it suitable for computing information rates. Sec.~\ref{sec:rates} explains how to compute lower bounds on capacity. Sec.~\ref{sec:mc} describes the multi-carrier approach with FDPA and Sec.~\ref{sec:results} provides numerical capacity lower bounds. Sec.~\ref{sec:conclusions} concludes the paper.

\section{Preliminaries}
\label{sec:prelim}
\subsection{Notation}
We use similar notation as in~\cite{garcia2020mismatched}. For instance, the Fourier transform of a function $u(t)$ is
\begin{equation}
\mathcal{F}\left(u(t)\right)=\mathcal{F}\left(u(t)\right)(\Omega)=\int_{-\infty}^{\infty}u(t)e^{-j\Omega t}\;\mathrm{d}t
\end{equation}
and the inverse Fourier transform of $U(\Omega)$ is $\mathcal{F}^{-1}\left(U(\Omega)\right)=\mathcal{F}^{-1}\left(U(\Omega)\right)(t)$. The \emph{dispersion operator} $\mathcal{D}_z$ is defined as
\begin{equation}
\mathcal{D}_z u(t)=\mathcal{F}^{-1}\left(e^{j\frac{\beta_2}{2}\Omega^2z}\mathcal{F}\left(u(t)\right)\right)
\end{equation}
which is the same as a convolution with an all-pass filter with the spectrum $e^{j\frac{\beta_2}{2}\Omega^2z}$. 

Fourier transforms, dispersion operators, and convolutions are linear operators $L$ for which we write
\begin{align}
  Lu(t) & = \int_{-\infty}^{\infty}u(\tau)K(\tau,t)\;\mathrm{d}\tau \label{eq:linear_operator1} \\
  L^*v(t) & = \int_{-\infty}^{\infty}v(\tau)K(t,\tau)^*\;\mathrm{d}\tau \label{eq:linear_operator2}
\end{align}
where $K$ is the kernel function and $L^*$ is the adjoint of $L$. We have the inner-product property 
\begin{equation}
  \int_{-\infty}^{\infty}Lu(t)\,v(t)^*\;\mathrm{d}t = \int_{-\infty}^{\infty}u(t)\,(L^*v(t))^*\;\mathrm{d}t.
  \label{eq:inner-product-property}
\end{equation}

A two-polarization signal is written as a vector $\mathbf{u}=(u, \overline{u})^T$. For vectors, the operators $\mathcal{F}$, $\mathcal{F}^{-1}$ and $\mathcal{D}_z$ are applied entrywise to the components. We consider unit-energy sinc-pulses
\begin{align}
    s(t) = \frac {1}{\sqrt{T}}\sinc\left(\frac{t}{T}\right)
\end{align}
where $\sinc(x)\defeq\sin(\pi x)/(\pi x)$ and $T$ is the symbol period.

\subsection{Dual-Polarization Propagation Model}
Consider an optical fiber of length $\mathcal{L}$. The polarization state of a signal changes randomly along the fiber due to randomly varying birefringence. If the change is fast enough, then the propagation of a 2-pol signal $\mathbf{u}(z, t)$ is described by the Manakov equation~\cite{manakov1973theory, wai1996polarization}:
\begin{equation}
\frac{\partial}{\partial z}\mathbf{u}=-j\frac{\beta_2}{2}\frac{\partial^2}{\partial t^2}\mathbf{u}+j\gamma f(z)\left\|\mathbf{u}\right\|^2\mathbf{u}+\frac{1}{\sqrt{f(z)}}\mathbf{n}
\label{eq:nlse}
\end{equation}
where $z$ is distance, $t$ is time, $\beta_2$ is the dispersion coefficient, and $\gamma=(8/9)2\pi n_2/(\lambda A_{\textrm{eff}})$ is the nonlinear coefficient. $A_{\textrm{eff}}$ is the effective area of the fiber, $n_2$ is the nonlinear index coefficient, $\lambda$ is the transmission wavelength, and the factor $8/9$ is due to the randomly varying birefringence. The scalar function $f(z)$ models attenuation and amplification along the fiber, and ideal distributed amplification (IDA) has $f(z)=1$. The ASE noise vector $\mathbf{n}(z, t)=(n(z,t),\overline{n}(z,t))^T$ has entries that are independent Wiener processes in $z$ such that, in the absence of signal ($\mathbf{u}=\mathbf{0}$) and nonlinearity ($\gamma=0$), the accumulated noise at a receiver of bandwidth $\mathcal{B}_{\textrm{ASE}}$ at $z=\mathcal{L}$ has two independent components that are Gaussian processes with autocorrelation function $N_{\textrm{ASE}}\mathcal{B}_{\textrm{ASE}}\mathrm{sinc}\left(\mathcal{B}_{\textrm{ASE}}(t-t')\right)$.

\section{Dual-Polarization RP Models}\label{sec:rp}

This section develops simplified models for the Manakov equation. We first derive the continuous-time RP solution of \eqref{eq:nlse}. We then consider WDM signaling and develop discrete-time models for dispersion compensation and digital back-propagation (DBP).

RP first solves the Manakov equation without the nonlinear term to obtain $\mathbf{u}_0(z,t)$ in~\eqref{eq:u0} and then treats the nonlinearity as a small additive term that depends on the linear solution, see~\eqref{eq:nlse_gamma_1}. As the nonlinear term is cubic and the receiver applies a matched filter, the effect of the nonlinearity depends on the integral over distance and time of the product of four copies of the (time-broadened) base pulse $s(z,t)=\mathcal{D}_{z} s(t)$, see~\eqref{eq:Ankk}. A graphical representation of these four-pulse interactions is given in~\cite{dar2016pulse}. The interactions create additive nonlinear interference (NLI) terms that are proportional to the products of three symbols, see~\eqref{eq:nli_sum}.

\subsection{Continuous-Time RP Model}
\label{subsec:CT-RP}
Similar to~\cite{mecozzi2012nonlinear}, we expand the signal $\mathbf{u}$ in powers of $\gamma$ which is assumed small:
\begin{equation}
\mathbf{u}(z, t)=\mathbf{u}_0(z, t)+\gamma \Delta\mathbf{u}(z, t)+\mathcal{O}(\gamma^2).
\label{eq:perturbation}
\end{equation}
We substitute~\eqref{eq:perturbation} in~\eqref{eq:nlse}, and solve the equations for the zeroth and first powers of $\gamma$. The zeroth-order equation is
\begin{equation}
\frac{\partial}{\partial z}\mathbf{u}_0(z, t)=-j\frac{\beta_2}{2}\frac{\partial^2}{\partial t^2}\mathbf{u}_0(z, t)+\frac{1}{\sqrt{f(z)}}\mathbf{n}(z, t).
\label{eq:zero-order-equation}
\end{equation}
Using the general expression
\begin{equation}
\frac{\partial}{\partial z}\left(\mathcal{D}_z g(z, t)\right)=-j\frac{\beta_2}{2}\frac{\partial^2}{\partial t^2}\left(\mathcal{D}_z g(z, t)\right)+\mathcal{D}_z\left(\frac{\partial}{\partial z}g(z, t)\right)
\label{eq:partial-z-expression}
\end{equation}
and choosing
\begin{equation}
g(z,t)=\mathbf{u}(0, t) + \int_{0}^{z}\mathcal{D}_{-z'}\frac{\mathbf{n}(z', t)}{f(z')}\;\mathrm{d}z'
\end{equation}
one can verify that the solution of \eqref{eq:zero-order-equation} is 
\begin{equation}
\mathbf{u}_0(z, t)=\mathbf{u}_{\textrm{LIN}}(z, t)+\mathbf{u}_{\textrm{ASE}}(z, t)
\label{eq:u0}
\end{equation}
where $\mathbf{u}_{\textrm{LIN}}(z, t)=\mathcal{D}_{z}\mathbf{u}(0, t)$ and
\begin{equation}
\mathbf{u}_{\textrm{ASE}}(z, t)=\mathcal{D}_z\left(\int_{0}^{z}\mathcal{D}_{-z'}\frac{\mathbf{n}(z', t)}{f(z')}\;\mathrm{d}z'\right). 
\end{equation}
This is similar to the 1-pol case and the properties derived in~\cite{mecozzi2012nonlinear} apply. The entries $u_{\textrm{ASE}}(z, t)$ and $\overline{u}_{\textrm{ASE}}(z, t)$ of $\mathbf{u}_{\textrm{ASE}}$ are independent, and their autocorrelation functions are
\begin{multline}
\left\langle u_{\textrm{ASE}}(z, t)\,u_{\textrm{ASE}}^*(z', t')\right\rangle \\ =\frac{N_{\textrm{ASE}}}{\kappa(\mathcal{L}, \mathcal{L})}\kappa(z, z')\mathcal{B}_{\textrm{ASE}}\mathrm{sinc}\left(\mathcal{B}_{\textrm{ASE}}(t-t')\right)
\label{eq:autocorr_w}
\end{multline}
where
\begin{equation}
	\kappa(z, z')=\int_{0}^{\min\{z,z'\}}\frac{1}{f(z'')}\;\mathrm{d}z''.
\end{equation}

The first-order equation in $\gamma$ is
\begin{equation}
\frac{\partial}{\partial z}\Delta\mathbf{u}=-j\frac{\beta_2}{2}\frac{\partial^2}{\partial t^2}\Delta\mathbf{u}+j f(z)\left\|\mathbf{u}_0\right\|^2\mathbf{u}_0 .
\label{eq:nlse_gamma_1}
\end{equation}
Again applying \eqref{eq:partial-z-expression}, the solution of \eqref{eq:nlse_gamma_1} is
\begin{multline}
\Delta \mathbf{u}(z,t) \\ = j\mathcal{D}_z\left(\int_0^{z}f(z')\mathcal{D}_{-z'}\left(\left\|\mathbf{u}_0(z', t)\right\|^2\mathbf{u}_0(z', t)\right)\;\mathrm{d}z'\right).
\label{eq:u_NL}
\end{multline}
The RP solution of the Manakov equation is thus
\begin{equation}
\mathbf{u}(z, t)=\mathbf{u}_{\textrm{LIN}}(z, t)+\mathbf{u}_{\textrm{ASE}}(z, t)+\mathbf{u}_{\textrm{NL}}(z, t)
\label{eq:rp_continuous}
\end{equation}
where $\mathbf{u}_{\textrm{NL}}(z, t)=\gamma \Delta \mathbf{u}_0(z,t)$.
The linear and ASE terms are similar to those in the scalar case~\cite{mecozzi2012nonlinear, garcia2020mismatched} but $\mathbf{u}_{\textrm{NL}}(z,t)$ couples the polarizations via $\left\|\mathbf{u}_0(z', t)\right\|^2$ in \eqref{eq:u_NL}.

\subsection{WDM with PAM}
\label{subsec:WDM}
We consider pulse amplitude modulation (PAM) with WDM. The WDM channels have indexes $c$ satisfying
\begin{equation}
c\in\mathcal{C}=\left\{c_{\min},\ldots,0,\ldots,c_{\max}\right\}
\end{equation}
where $c_{\min}\le 0$ and $c_{\max}\ge 0$.
The center angular frequency of channel $c$ is $\Omega^{(c)}$, $c\in\mathcal{C}$, and we choose $\Omega^{(c)}=0$. The two components of the transmitted signal are
\begin{align}
u(0, t)= & \sum_{m=-\infty}^{\infty}x_{m}s\left(t-mT\right) \nonumber \\
& +\sum_{c\ne 0}e^{j\Omega^{(c)} t}\sum_{k=-\infty}^{\infty}b_{k}^{(c)}s\left(t-kT-\Delta T^{(c)}\right) \label{eq:wdm1} \\
\overline{u}(0, t)= & \sum_{m=-\infty}^{\infty}\overline{x}_{m}s\left(t-mT-\overline{\Delta T}\right) \nonumber \\
& +\sum_{c\ne 0}e^{j\Omega^{(c)} t}\sum_{k=-\infty}^{\infty}\overline{b}_{k}^{(c)}s\left(t-kT-\overline{\Delta T}^{(c)}\right)
\label{eq:wdm2}
\end{align}
where $T$ is the symbol period and $s(t)$ is the base pulse that is taken to be the same for all channels. The delays $\overline{\Delta T}$, $\Delta T^{(c)}$ and $\overline{\Delta T}^{(c)}$ allow asynchronous transmission. These delays are here measured with respect to the first polarization of the channel of interest (COI) with $c=0$, i.e., $\Delta T=\Delta T^{(0)}=0$. The symbols transmitted in the COI are $x_{m}$ for the first polarization and $\overline{x}_{m}$ for the second. The symbols transmitted in the interfering channels are $b_{k}^{(c)}$ and $\overline{b}_{k}^{(c)}$.

The base pulse is chosen as a root-Nyquist pulse with unit energy, $\|s(t)\|^2=1$, with most of its energy in the frequency band $|\Omega|\le\pi\mathcal{B}$. We assume $2\pi\mathcal{B}\le\min_c\left(\Omega^{(c+1)}-\Omega^{(c)}\right)$ so that the channels can be separated in frequency. The transmitted symbol sequences $\left\{ X_{m}\right\}$, $\left\{ \overline{X}_{m}\right\}$, $\{ B_{k}^{(c)}\}$ and $\{ \overline{B}_{k}^{(c)}\}$ are independent and identically distributed (i.i.d.) proper complex processes. The energies are $\left\langle|X_{m}|^2\right\rangle=E$ and $\left\langle|B_{k}^{(c)}|^2\right\rangle=E^{(c)}$, and similarly for $\overline{E}$ and $\overline{E}^{(c)}$. The launch power of the first polarization of the COI is thus $\mathcal{P}=E/T$
. We define the fourth moments $Q^{(c)}=\left\langle|B_{k}^{(c)}|^4\right\rangle$ and $\overline{Q}^{(c)}=\left\langle|\overline{B}_{k}^{(c)}|^4\right\rangle$. 

We focus on the received symbols in the first polarization. All results apply to the second polarization by letting the variables with ``bar'' denote the first polarization and the variables without ``bar'' denote the second.

The receiver uses a band-pass filter $h_{\mathcal{B}}(t)$ to isolate the COI. We assume that $h_{\mathcal{B}}(t)*s(t)=s(t)$. The receiver then applies either dispersion compensation $\mathcal{D}_{-\mathcal{L}}$ or 2-pol digital back-propagation (DBP) to the COI, followed by matched filtering and sampling for each polarization. We proceed to develop discrete-time models for both cases.

\subsection{Discrete-Time Model for Dispersion Compensation}
\label{subsec:DT-DC}
For dispersion compensation the sampled symbols of the first polarization of the COI are the inner products
\begin{equation}
y_{m}=\int_{-\infty}^{\infty}s\left(t-mT\right)^*\left\{\mathcal{D}_{-\mathcal{L}}\left[h_{\mathcal{B}}(t)*u(\mathcal{L}, t)\right]\right\}\diff t.
\label{eq:mf}
\end{equation}
where $*$ denotes convolution. Inserting~\eqref{eq:rp_continuous} into~\eqref{eq:mf}, we have
\begin{equation}
y_{m}=x_{m}+w_{m}+\Delta x_{m}
\label{eq:y_1}
\end{equation}
where the linear noise term is
\begin{align}
w_{m} =\int_{-\infty}^{\infty} & s\left(t-mT\right)^*
\left\{\mathcal{D}_{-\mathcal{L}}\left[h_{\mathcal{B}}(t)*u_{\mathrm{ASE}}(\mathcal{L}, t)\right]\right\}\diff t.
\end{align}
The $\mathcal{D}_{-\mathcal{L}}$ operator does not change the statistical properties of the noise~\cite{mecozzi2012nonlinear}. If most of the energy of $s(t)$ is contained in the COI with bandwidth $\mathcal{B}$, then the $\left\{ W_{m}\right\}$ are i.i.d. proper complex Gaussian with
\begin{equation}
\left\langle W_{m}W_{m'}^*\right\rangle=N_{\textrm{ASE}}\delta[m'-m]
\end{equation}
where $\delta[\ell]=1$ if $\ell=0$ and $\delta[\ell]=0$ otherwise. The $W_{m}$ are uncorrelated with the $\overline{W}_{m}$.

The NLI term $\Delta x_{m}$ is
\begin{equation}
\Delta x_{m}=\int_{-\infty}^{\infty}s\left(t-mT\right)^*\left\{\mathcal{D}_{-\mathcal{L}}\left[h_{\mathcal{B}}(t)*u_{\textrm{NL}}(\mathcal{L}, t)\right]\right\}\diff t.
\label{eq:nli}
\end{equation}
As cross-phase modulation (XPM) limits the capacity~\cite{essiambre_limits}, we neglect signal-noise mixing by replacing $\mathbf{u}_0$ with $\mathbf{u}_{\textrm{LIN}}$ in~\eqref{eq:u_NL}. Substituting the resulting $u_{\textrm{NL}}(\mathcal{L},t)$ into~\eqref{eq:nli} and using the linearity of the $\mathcal{D}_z$ and $*$ operators, the dispersion operators $\mathcal{D}_{-\mathcal{L}}$ and $\mathcal{D}_{\mathcal{L}}$ cancel, and \eqref{eq:inner-product-property} gives
\begin{align}
\int_{-\infty}^{\infty} & s(t)^*\left\{h_{\mathcal{B}}(t)*\left[\mathcal{D}_{-z} v(t)\right]\right\}\diff t \nonumber \\
& =\int_{-\infty}^{\infty}\left\{\mathcal{D}_{z}\left[h_{\mathcal{B}}(-t)^**s(t)\right]\right\}^* v(t)\diff t.
\end{align}

We now assume that $h_{\mathcal{B}}(-t)^*=h_{\mathcal{B}}(t)$, and use $h_{\mathcal{B}}(t) * s(t)=s(t)$. We write $s(z,t)=\mathcal{D}_{z}s(t)$ to obtain
\begin{align}
\Delta x_{m} & =j\gamma\int_{0}^{\mathcal{L}}f(z)\int_{-\infty}^{\infty} s(z, t-mT)^* \nonumber \\
& \cdot\left(\left|u_{\textrm{LIN}}(z, t)\right|^2+\left|\overline{u}_{\textrm{LIN}}(z, t)\right|^2\right)u_{\textrm{LIN}}(z, t)\diff t \diff z.
\label{eq:Delta_x}
\end{align}
We next insert the WDM signals~\eqref{eq:wdm1} and~\eqref{eq:wdm2} into~\eqref{eq:Delta_x} and use the following identity for a delayed and frequency-shifted pulse $s(t)$:
\begin{multline}
\mathcal{D}_z\left[e^{j\Omega^{(c)}t}s(t-t_0)\right] \\ =e^{j\frac{\beta_2}{2}(\Omega^{(c)})^2z}e^{j\Omega^{(c)}t}s(z, t-t_0+\beta_2\Omega^{(c)}z).
\label{eq:group_delay}
\end{multline}
Ignoring the four-wave mixing (FWM) terms, we obtain
\begin{align}
\Delta x_{m} & =j\sum_{\substack{n\\ k,k'}}S_{n,k,k'}x_{ n+m}x_{k+m}x_{k'+m}^* \nonumber \\
&  +j\sum_{\substack{n\\ k,k'}}\tilde{S}_{n,k,k'}x_{n+m}\overline{x}_{k+m}\overline{x}_{k'+m}^* \nonumber \\
& +j\sum_{c\ne 0}\sum_{\substack{n\\ k,k'}}C_{n,k,k'}^{(c)}x_{n+m}b_{k+m}^{(c)}b_{k'+m}^{(c), *} \nonumber \\
& +j\sum_{c\ne 0}\sum_{\substack{n\\ k,k'}}\tilde{C}_{n,k,k'}^{(c)}x_{n+m}\overline{b}_{k+m}^{(c)}\overline{b}_{k'+m}^{(c), *} \nonumber \\
& +j\sum_{c\ne 0}\sum_{\substack{n\\ k,k'}}D_{n,k,k'}^{(c)}\overline{x}_{n+m}b_{k+m}^{(c)}\overline{b}_{k'+m}^{(c),*}.
\label{eq:nli_sum}
\end{align}
We explain the terms of~\eqref{eq:nli_sum} in Sec.~\ref{subsec:nli_coefficients} below.

\subsection{Discrete-Time Model for DBP}
\label{subsec:DT-DBP}
For DBP the received signal $\mathbf{u}(\mathcal{L},t)$ in \eqref{eq:rp_continuous} is filtered with $h_{\mathcal{B}}(t)$ and propagated for distance $\mathcal{L}$ with $\beta_2$ and $\gamma$ replaced by $-\beta_2$ and $-\gamma$, respectively, and $f(z)$ replaced by $f(\mathcal{L}-z)$. Again using RP, the DBP signal at distance $z$ (or overall distance $\mathcal{L}+z$) is as in \eqref{eq:rp_continuous} but without the noise term:
\begin{equation}
  \mathbf{u}_{\textrm{DBP}}(z,t)=\tilde{\mathbf{u}}_{\textrm{LIN}}(z, t)+\tilde{\mathbf{u}}_{\textrm{NL}}(z,t)
  \label{eq:u_DBP}
\end{equation}
where
\begin{equation}
  \tilde{\mathbf{u}}_{\textrm{LIN}}(z,t) = \mathcal{D}_{-z}\left[h_{\mathcal{B}}(t)*\mathbf{u}(\mathcal{L}, t)\right]
  \label{eq:ut_LIN}
\end{equation}
is the dispersion compensation output
and
\begin{multline}
  \tilde{\mathbf{u}}_{\textrm{NL}}(z,t) = -j\gamma \, \mathcal{D}_{-z}\left(\int_0^{z}f(\mathcal{L}-z') \right. \\ \left. \cdot \mathcal{D}_{z'}\left(\left\|\tilde{\mathbf{u}}_{\textrm{LIN}}(z', t)\right\|^2 \tilde{\mathbf{u}}_{\textrm{LIN}}(z', t)\right)\;\mathrm{d}z'\right).
\label{eq:ut_NL}
\end{multline}
is a nonlinear correction term (see \eqref{eq:u_NL} and \eqref{eq:rp_continuous}).

We proceed to simplify \eqref{eq:ut_NL}. Recall from  \eqref{eq:rp_continuous} that 
\begin{equation}
  \mathbf{u}(\mathcal{L},t) = \mathcal{D}_{\mathcal{L}}\mathbf{u}(0,t)+\mathbf{u}_{\textrm{ASE}}(\mathcal{L},t)+\mathbf{u}_{\textrm{NL}}(\mathcal{L},t).
\end{equation}
We now modify $\tilde{\mathbf{u}}_{\textrm{LIN}}(z', t)$ in \eqref{eq:ut_NL} by neglecting the ASE term $\mathbf{u}_{\textrm{ASE}}(\mathcal{L},t)$ and the NLI term $\mathbf{u}_{\textrm{NL}}(\mathcal{L},t)$. One may justify removing the ASE term as neglecting signal-noise mixing, and the NLI term because it has $\gamma^2$ or smaller terms only. The correction \eqref{eq:ut_NL} at $z=\mathcal{L}$ is thus
\begin{multline}
   -j\gamma \int_0^{\mathcal{L}}f(z)\,\mathcal{D}_{-z}\left(\left\|\mathcal{D}_z \mathbf{u}^{(0)}(0, t)\right\|^2 \mathcal{D}_z \mathbf{u}^{(0)}(0, t)\right)\;\mathrm{d}z
\label{eq:ut_NL2}
\end{multline}
where we have applied a change of variables to the integration and written the COI launch signal as $\mathbf{u}^{(0)}(0,t)=h_{\mathcal{B}}(t)*\mathbf{u}(0, t)$. The correction term \eqref{eq:ut_NL2} followed by matched filtering and sampling gives \eqref{eq:Delta_x} but with negative $\gamma$ and where $\mathbf{u}_{\textrm{LIN}}(z,t)$ has only the contributions of the COI. DBP thus gives \eqref{eq:nli_sum} but without the sums with $S_{n,k,k'}$ and $\tilde{S}_{n,k,k'}$.
In other words, the analysis confirms the intuition that, under RP approximation, DBP exactly compensates the intra-channel nonlinearity and does not change inter-channel nonlinearities with respect to dispersion compensation.

\subsection{Nonlinear Coefficients}
\label{subsec:nli_coefficients}
The NLI coefficients in~\eqref{eq:nli_sum} can be expressed using the following general form:
\begin{align}
    & A_{n,k,k'}(t_1,t_2,t_3)  =\gamma\int_0^{\mathcal{L}}f(z)\int_{-\infty}^{\infty}s(z, t)^*s(z, t-nT-t_1) \nonumber \\ & \qquad \cdot s(z, t-kT-t_2) s(z, t-k'T-t_3)^*\diff t \diff z. \label{eq:Ankk}
\end{align}

\subsubsection{Self-Phase Modulation (SPM)}
\begin{figure}[tbp]\centering
	\setlength{\figurewidth}{0.9\linewidth}
	\setlength{\figureheight}{0.475\figurewidth}
	\input{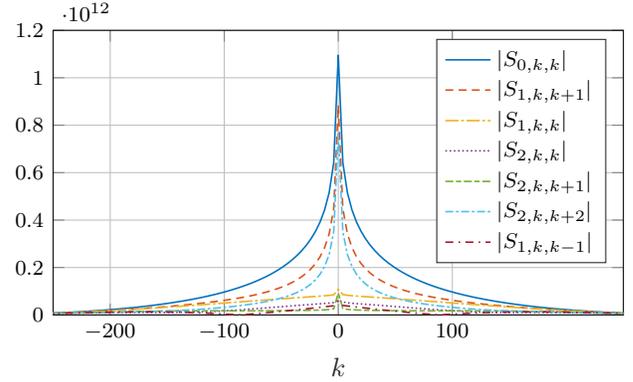}%
	\caption{SPM coefficient magnitudes $|S_{n,k,k'}|$ for a 1000-km link with the parameters in Table~\ref{tab:parameters} in the time-synchronized case, where $S_{n,k,k'}=\tilde{S}_{n,k,k'}$.}
	\label{fig:Snkk}
\end{figure}

As shown above, the SPM terms are present if the receiver uses dispersion compensation but not DBP on the COI. The \emph{self-polarization SPM} terms represent interactions of four pulses from the first polarization (1st-pol) of the COI:
\begin{equation}
jS_{n,k,k'}x_{n+m}x_{k+m}x_{k'+m}^* ;
\label{eq:S_terms}
\end{equation}
the \emph{cross-polarization SPM} terms represent interactions of two pulses from the 1st-pol of the COI and two pulses from the 2nd-pol of the COI:
\begin{equation}
j\tilde{S}_{n,k,k'}x_{n+m}\overline{x}_{k+m}\overline{x}_{k'+m}^* ;
\label{eq:S_cross_terms}
\end{equation}
the SPM coefficients are
\begin{align}
S_{n, k, k'} & = A_{n,k,k'}(0,0,0) \label{eq:Snkk} \\
\tilde{S}_{n, k, k'} & = A_{n,k,k'}(0,\overline{\Delta T},\overline{\Delta T}). \label{eq:Sbarnkk}
\end{align}
Fig.~\ref{fig:Snkk} shows the $|S_{n,k,k'}|$ for the parameters in Table~\ref{tab:parameters} and with synchronized polarizations ($\overline{\Delta T}=0$). As shown in~\cite{mecozzi2000analysis}, the coefficients with $k'=n+k$ are larger than the others.

\begin{table}[tbp]\centering
	\caption{System parameters}
	\label{tab:parameters}
	\begin{tabular}{ccc}
		\hline
		\textbf{Parameter} & \textbf{Symbol} & \textbf{Value} \\
		\hline
		Attenuation coefficient & $\alpha$ & $0.2\;\textrm{dB}/\mathrm{km}$ \\
		Dispersion coefficient & $\beta_2$ & $-21.7\;\mathrm{ps}^2/\mathrm{km}$ \\
		Nonlinear coefficient & $\gamma$ & $1.27\;\mathrm{W}^{-1}\mathrm{km}^{-1}$  \\
		Phonon occupancy factor & $\eta$ & $1$  \\
		Transmit pulse shape & $s(t)$ & sinc \\
		Channel bandwidth & $\mathcal{B}$ & $50\;\textrm{GHz}$ \\
		Channel spacing & $\Omega^{(1)}/(2\pi)$ & $50\;\textrm{GHz}$ \\
		\hline
	\end{tabular}
\end{table}

\subsubsection{Cross-Phase Modulation (XPM)}
XPM gives the strongest NLI terms at the capacity peak for systems with DBP of the COI. For all $c\ne 0$, the \emph{self-polarization XPM} terms represent interactions of two pulses from the 1st-pol of the COI and two pulses from the 1st-pol of an interfering channel (IC):
\begin{equation}
jC_{n,k,k'}^{(c)}x_{n+m}b_{k+m}^{(c)}b_{k'+m}^{(c), *} ;
\end{equation}
the \emph{cross-polarization XPM} terms represent interactions of two pulses from the 1st-pol of the COI and two pulses from the 2nd-pol of an IC:
\begin{equation}
j\tilde{C}_{n,k,k'}^{(c)}x_{n+m}\overline{b}_{k+m}^{(c)}\overline{b}_{k'+m}^{(c), *} ;
\end{equation}
and the \emph{mixed-polarization XPM} terms represent interactions of one pulse from the 1st-pol of the COI, one from the 2nd-pol of the COI, one from the 1st-pol of an IC, and one from the 2nd-pol of an IC:
\begin{equation}
jD_{n,k,k'}^{(c)}\overline{x}_{n+m}b_{k+m}^{(c)}\overline{b}_{k'+m}^{(c),*} ;
\end{equation}
the XPM coefficients are
\begin{align}
& C_{n,k,k'}^{(c)} = 2 A_{n,k,k'}\left(0,\Delta T^{(c)}-\beta_2\Omega^{(c)}z,\Delta T^{(c)}-\beta_2\Omega^{(c)}z \right) \label{eq:Cnkk} \\
& \tilde{C}_{n,k,k'}^{(c)} = A_{n,k,k'}\left(0,\overline{\Delta T}^{(c)}-\beta_2\Omega^{(c)}z,\overline{\Delta T}^{(c)}-\beta_2\Omega^{(c)}z \right)
\label{eq:Cbarnkk} \\
& D_{n,k,k'} \nonumber \\
& \quad = A_{n,k,k'}\left(\overline{\Delta T},\Delta T^{(c)}-\beta_2\Omega^{(c)}z,\overline{\Delta T}^{(c)}-\beta_2\Omega^{(c)}z \right). \label{eq:Dnkk}
\end{align}
All other terms involving only channels $0$ and $c$, such as $b_{k'}^{(c)}\overline{b}_{k}^{(c)}\overline{x}_{k'}^*$ vanish because the $e^{j\Omega^{(c)}t}$ in~\eqref{eq:Delta_x} do not cancel, and the terms fall outside the receiver filter bandwidth.

\begin{figure}[tbp]
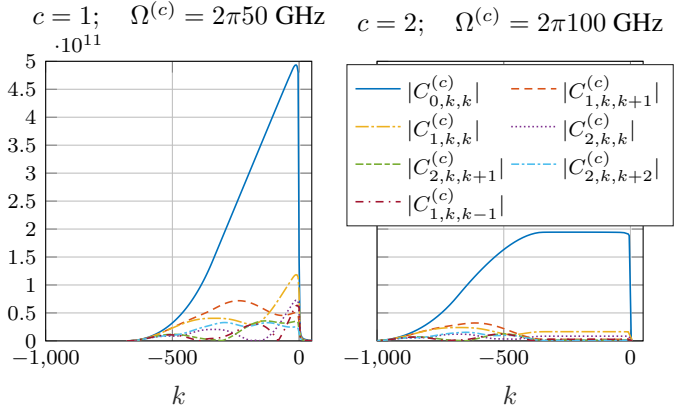
\centering
	\setlength{\figurewidth}{0.42\linewidth}
	\setlength{\figureheight}{1\figurewidth}
	\input{Graphs/Cnkk_new.tex}%
	\input{Graphs/Cnkk2_new.tex}
	\caption{XPM coefficient magnitudes $|C_{n,k,k'}^{(c)}|$ for a 1000-km link with the parameters in Table~\ref{tab:parameters} for the time-synchronized case, where $C_{n,k,k'}^{(c)}=2\tilde{C}_{n,k,k'}^{(c)}=2D_{n,k,k'}^{(c)}$.}
	\label{fig:Cnkk}
\end{figure}

We focus on the synchronized case ($\Delta T^{(c)}=\overline{\Delta T}^{(c)}$
for all $c$) which gives the highest rates in our simulations. We then have $C_{n,k,k'}^{(c)}=2\tilde{C}_{n,k,k'}^{(c)}=2D_{n,k,k'}^{(c)}$. Fig.~\ref{fig:Cnkk} shows that the NLI from channel $c=2$ has smaller magnitude but longer memory than the NLI from channel $c=1$.

In the synchronized case, we have
\begin{equation}
D_{n,k,k'}^{(c)}=D_{-n, k'-n, k-n}^{(c), *}.
\label{eq:conj_symmetry}
\end{equation}
If $s(t)=s(-t)$, we additionally have
\begin{equation}
D_{n,k,k'}^{(c)} =D_{k'-k,k'-n,k'}^{(c)}. \label{eq:other_symmetry} 
\end{equation}
If $s(t)=s(-t)$ and $\Omega^{(-c)}=-\Omega^{(c)}$, we also have
\begin{equation}
D_{n,k,k'}^{(c)}=D_{-n, -k, -k'}^{(-c)}. \label{eq:neg_symmetry}
\end{equation}
Equations~\eqref{eq:conj_symmetry}-\eqref{eq:neg_symmetry} similarly hold for the other NLI coefficients $S_{n,k,k'}, \tilde{S}_{n,k,k'}, C_{n,k,k'}$ and $ \tilde{C}_{n,k,k'}$.

\section{2pCPAN Model}\label{sec:2pcpan}
We consider DBP on the COI and thus neglect the SPM terms. As for the 1-pol case in~\cite{garcia2020mismatched}, we separate~\eqref{eq:nli_sum} into terms that depend on the current 2-pol symbol $\mathbf{x}_m=(x_m,\overline{x}_m)^T$ and those that do not:
\begin{equation}
\left(\begin{matrix}
\Delta x_{m} \\ \overline{\Delta x}_{m}
\end{matrix}\right) = j\left(\begin{matrix}
\theta_{m} & \psi_{m} \\
\overline{\psi}_{m} & \overline{\theta}_{m}
\end{matrix}\right)\left(\begin{matrix}
x_{m} \\ \overline{x}_{m}
\end{matrix}\right)+\left(\begin{matrix}
v_{m} \\ \overline{v}_{m}
\end{matrix}\right)
\label{eq:nli_matrix_1}
\end{equation}
where we have defined
\begin{align}
\theta_{m} & =\sum_{c\ne 0}\sum_{ k,k'}C_{0,k,k'}^{(c)}b_{k+m}^{(c)}b_{k'+m}^{(c), *} \nonumber \\
& \quad +\sum_{c\ne 0}\sum_{ k,k'}\tilde{C}_{0,k,k'}^{(c)}\overline{b}_{k+m}^{(c)}\overline{b}_{k'+m}^{(c), *} \label{eq:theta}
\\
\psi_{m} & =\sum_{c\ne 0}\sum_{ k,k'}D_{0,k,k'}^{(c)}b_{k+m}^{(c)}\overline{b}_{k'+m}^{(c),*} 
\label{eq:psi_im}
\\
v_{m} & = j\sum_{c\ne 0}\sum_{\substack{n\ne 0 \\ k,k'}}C_{n,k,k'}^{(c)}x_{n+m}b_{k+m}^{(c)}b_{k'+m}^{(c), *} \nonumber \\
& \quad + j\sum_{c\ne 0}\sum_{\substack{n\ne 0 \\ k,k'}}\tilde{C}_{n,k,k'}^{(c)}x_{ n+m}\overline{b}_{k+m}^{(c)}\overline{b}_{k'+m}^{(c), *} \nonumber \\
& \quad + j\sum_{c\ne 0}\sum_{\substack{n\ne 0 \\ k,k'}}D_{n,k,k'}^{(c)}\overline{x}_{ n+m}b_{k+m}^{(c)}\overline{b}_{k'+m}^{(c),*}
\label{eq:v}
\end{align}
and $\overline{\theta}_m$, $\overline{\psi}_m$, and $\overline{v}_m$ are obtained by swapping $x_\ell$ with $\overline{x}_\ell$ and $b_\ell^{(c)}$ with $\overline{b}_\ell^{(c)}$ in~\eqref{eq:theta}-\eqref{eq:v}. From~\eqref{eq:Cnkk} and~\eqref{eq:Cbarnkk}, we have $C_{0,k,k'}^{(c)}=C_{0,k',k}^{(c),*}$ and $\tilde{C}_{0,k,k'}^{(c)}=\tilde{C}_{0,k',k}^{(c),*}$. Therefore, $\theta_{m}$ and $\overline{\theta}_m$ are real.

With $\Delta T^{(c)}=\overline{\Delta T}^{(c)}$ for all $c$, from~\eqref{eq:conj_symmetry} we have $D_{0,k,k'}^{(c)}=D_{0,k',k}^{(c), *}$ and therefore
\begin{equation}
\overline{\psi}_{m} = \psi_{m}^*.
\end{equation}

We now apply the following approximation valid to first order in $\gamma$:
\begin{align}
\mathbf{I}+j\left(\begin{matrix}
\theta_{m} & \psi_{m} \\
\psi_{m}^* & \overline{\theta}_{m}
\end{matrix}\right) \approx \underbrace{\exp\left[j\left(\begin{matrix}
\theta_{m} & \psi_{m} \\
\psi_{m}^* & \overline{\theta}_{m}
\end{matrix}\right)\right]}_{\mathbf{M}_m}.
\label{eq:M}
\end{align}
The motivation for the approximation is that both the analytical form of the Manakov equation and numerical results suggest a unitary rotation in two-dimensional (2D) complex space. Note that $\mathbf{M}_m$ in \eqref{eq:M} has the general form of such a rotation where the matrix argument of the exponential is skew-Hermitian. A unitary rotation also simplifies computing the output entropy in Section~\ref{sec:h_Y}. Substituting~\eqref{eq:nli_matrix_1} in~\eqref{eq:y_1}, and using~\eqref{eq:M}, we obtain the 2pCPAN model of the 2-pol optical channel:
\begin{equation}
\left(\begin{matrix}y_{m} \\ \overline{y}_{m}\end{matrix}\right)=\mathbf{M}_m\left(\begin{matrix}x_{m} \\ \overline{x}_{m}\end{matrix}\right)+\left(\begin{matrix}w_{m} \\ \overline{w}_{m}\end{matrix}\right)+\left(\begin{matrix}v_{m} \\ \overline{v}_{m}\end{matrix}\right)
\label{eq:cpan_model}
\end{equation}
where the $W_{m}$ and the $\overline{W}_m$ are i.i.d. proper complex Gaussian with variance $N_{\textrm{ASE}}$. The matrix $\mathbf{M}_m$ in~\eqref{eq:M} represents the NLI that causes a complex 2D rotation of the transmitted symbols, and the $v_{m}$ and $\overline{v}_m$ are the residual NLI that is not captured by $\mathbf{M}_m$. The $\theta_{m}$ and $\overline{\theta}_m$ in $\mathbf{M}_m$ represent phase noise in each polarization, and $\psi_m$ represents a unitary coupling between polarizations.

\subsection{First- and Second-Order Statistics}\label{sec:statistics}
We compute statistics of $\{\Theta_{m}\}$, $\{\Psi_m\}$, and $\{V_m\}$. The statistics of $\{\overline{\Theta}_{m}\}$ and $\{\overline{V}_m\}$ are respectively the same as those for $\{\Theta_{m}\}$ and $\{V_m\}$ after swapping $E^{(c)}$ and $Q^{(c)}$  with $\overline{E}^{(c)}$ and $\overline{Q}^{(c)}$, respectively.

\subsubsection{Means}
We have
\begin{align}
& \left\langle\Theta_{m}\right\rangle=\sum_{c\ne 0}E^{(c)}\sum_k C_{0,k,k}^{(c)}+\sum_{c\ne 0}\overline{E}^{(c)}\sum_k \tilde{C}_{0,k,k}^{(c)} \label{eq:E_theta} \\
& \left\langle\Psi_m\right\rangle=\left\langle V_{m}\right\rangle=0.
\label{eq:E_psi_v}
\end{align}

\subsubsection{Second-Order Statistics of $\Theta_{m}$}
The autocovariance function of $\{\Theta_{m}\}$ is
\begin{align}
r_{\Theta}[\ell]&\defeq\left\langle\Theta_{m}\Theta_{m+\ell}\right\rangle-\left\langle\Theta_{m}\right\rangle\left\langle\Theta_{m+\ell}\right\rangle \nonumber \\ 
& =\sum_{c\ne 0}\left(Q^{(c)}-\left(E^{(c)}\right)^2\right)\sum_k C_{0,k,k}^{(c)}C_{0,k-\ell,k-\ell}^{(c), *} \nonumber \\ 
& +\sum_{c\ne 0}\left(E^{(c)}\right)^2\sum_{k\ne k'} C_{0,k,k'}^{(c)}C_{0,k-\ell,k'-\ell}^{(c), *} \nonumber \\ 
& +\sum_{c\ne 0}\left(\overline{Q}^{(c)}-\left(\overline{E}^{(c)}\right)^2\right)\sum_k \tilde{C}_{0,k,k}^{(c)}\tilde{C}_{0,k-\ell,k-\ell}^{(c), *} \nonumber \\ 
& +\sum_{c\ne 0}\left(\overline{E}^{(c)}\right)^2\sum_{k\ne k'} \tilde{C}_{0,k,k'}^{(c)}\tilde{C}_{0,k-\ell,k'-\ell}^{(c), *}.
\label{eq:r_theta}
\end{align}
The crosscovariance function of $\{\Theta_{m}\}$ and $\{\overline{\Theta}_{m}\}$ is
\begin{align}
& \tilde{r}_{\Theta}[\ell]\defeq\left\langle\Theta_{m}\overline{\Theta}_{m+\ell}\right\rangle-\left\langle\Theta_{m}\right\rangle\left\langle\overline{\Theta}_{m+\ell}\right\rangle \nonumber \\ 
& =\sum_{c\ne 0}\left(Q^{(c)}-\left(E^{(c)}\right)^2\right)\sum_k C_{0,k,k}^{(c)}\tilde{C}_{0,k-\ell,k-\ell}^{(c), *} \nonumber \\ 
& +\sum_{c\ne 0}\left(E^{(c)}\right)^2\sum_{k\ne k'} C_{0,k,k'}^{(c)}\tilde{C}_{0,k-\ell,k'-\ell}^{(c), *} \nonumber \\ 
& +\sum_{c\ne 0}\left(\overline{Q}^{(c)}-\left(\overline{E}^{(c)}\right)^2\right)\sum_k \tilde{C}_{0,k,k}^{(c)}C_{0,k-\ell,k-\ell}^{(c), *} \nonumber \\ 
& +\sum_{c\ne 0}\left(\overline{E}^{(c)}\right)^2\sum_{k\ne k'} \tilde{C}_{0,k,k'}^{(c)}C_{0,k-\ell,k'-\ell}^{(c), *}.
\label{eq:r_bar_theta}
\end{align}
When $\Delta T^{(c)}=\overline{\Delta T}^{(c)}$ for all $c$, we have $\tilde{C}_{0,k,k'}^{(c)}=C_{0,k,k'}^{(c)}/2$. Furthermore, if $E^{(c)}=\overline{E}^{(c)}$ and $Q^{(c)}=\overline{Q}^{(c)}$ for all $c$, then~\eqref{eq:r_theta} and~\eqref{eq:r_bar_theta} simplify to
\begin{align}
r_{\Theta}[\ell] & = r_{\overline{\Theta}}[\ell] =\frac{5}{4}\tilde{r}_\Theta[\ell]  \label{eq:crosscorr_theta_frac} \\
\tilde{r}_{\Theta}[\ell] & =\sum_{c\ne 0} \left(Q^{(c)}-\left(E^{(c)}\right)^2\right)\sum_k C_{0,k,k}^{(c)}C_{0,k-\ell,k-\ell}^{(c), *} \nonumber \\ 
& + \left(E^{(c)}\right)^2\sum_{c\ne 0}\sum_{k\ne k'} C_{0,k,k'}^{(c)}C_{0,k-\ell,k'-\ell}^{(c), *}.
\label{eq:r_thetas_synchr}
\end{align}
The phase noises across polarizations are thus almost as strongly correlated as within each polarization.

\subsubsection{Second-Order Statistics of $\Psi_{m}$} 
The autocorrelation function of $\{\Psi_m\}$ is
\begin{align}
r_{\Psi}[\ell]\defeq \left\langle\Psi_{m}\Psi_{m+\ell}^*\right\rangle=\sum_{c\ne 0} E^{(c)} \overline{E}^{(c)} \sum_{k, k'} D_{0, k, k'}^{(c)}D_{0, k-\ell, k'-\ell}^{(c), *}.
\label{eq:r_psi}
\end{align}
If $\Delta T^{(c)}=\overline{\Delta T}^{(c)}$ for all $c$, then from~\eqref{eq:conj_symmetry} we have $D_{0,k,k'}^{(c)}=D_{0,k',k}^{(c),*}$ and $r_{\Psi}[\ell]$ is real. If we have Gaussian inputs and if $E^{(c)}=\overline{E}^{(c)}$ for all $c$, then $Q^{(c)}=2(E^{(c)})^2$ and
\begin{equation}
r_\Psi[\ell]=\frac{1}{5}r_{\Theta}[\ell].
\label{eq:r_psi_frac}
\end{equation}
From~\eqref{eq:psi_im}, the $\{\Psi_m\}$ are proper complex: for all $\ell$ we have
\begin{equation}
\left\langle\Psi_{m}\Psi_{m+\ell}\right\rangle=0.
\end{equation}

\subsubsection{Second-Order Statistics of $V_{m}$}
The autocorrelation function of $\{V_{m}\}$ is given by
\begin{align}
& r_{V}[\ell]\defeq\left\langle V_{m}V_{m+\ell}^*\right\rangle \nonumber \\
&
=E\sum_{c\ne 0}\left(Q^{(c)}-\left(E^{(c)}\right)^2\right)\sum_{\substack{n\ne 0 \\ n\ne\ell}} \sum_{k} C_{n,k,k}^{(c)}C_{n-\ell,k-\ell, k-\ell}^{(c), *} \nonumber \\
&
+E\sum_{c\ne 0}\left(E^{(c)}\right)^2\sum_{\substack{n\ne 0 \\ n\ne\ell}} \sum_{k\ne k'} C_{n,k,k'}^{(c)}C_{n-\ell,k-\ell, k'-\ell}^{(c), *} \nonumber \\
& +E\sum_{c\ne 0}\left(\overline{Q}^{(c)}-\left(\overline{E}^{(c)}\right)^2\right)\sum_{\substack{n\ne 0 \\ n\ne\ell}} \sum_{k} \tilde{C}_{n,k,k}^{(c)}\tilde{C}_{n-\ell,k-\ell, k-\ell}^{(c), *} \nonumber \\
&
+E\sum_{c\ne 0}\left(\overline{E}^{(c)}\right)^2\sum_{\substack{n\ne 0 \\ n\ne\ell}} \sum_{k\ne k'} \tilde{C}_{n,k,k'}^{(c)}\tilde{C}_{n-\ell,k-\ell, k'-\ell}^{(c), *} \nonumber \\
& + E\sum_{\substack{c\ne 0 \\ c'\ne 0}}E^{(c)}E^{(c')}\sum_{\substack{n\ne 0 \\ n\ne\ell}} \sum_{k,k'} C_{n,k,k}^{(c)}C_{n-\ell,k'-\ell, k'-\ell}^{(c'), *} \nonumber \\
&
+ E\sum_{\substack{c\ne 0 \\ c'\ne 0}}\overline{E}^{(c)}\overline{E}^{(c')}\sum_{\substack{n\ne 0 \\ n\ne\ell}} \sum_{k,k'} \tilde{C}_{n,k,k}^{(c)}\tilde{C}_{n-\ell,k'-\ell, k'-\ell}^{(c'), *} \nonumber \\
& + E\sum_{\substack{c\ne 0 \\ c'\ne 0}}E^{(c)}\overline{E}^{(c')}\sum_{\substack{n\ne 0 \\ n\ne\ell}} \sum_{k,k'} {C}_{n,k,k}^{(c)}\tilde{C}_{n-\ell,k'-\ell, k'-\ell}^{(c'), *} \nonumber \\
&
+ E\sum_{\substack{c\ne 0 \\ c'\ne 0}}\overline{E}^{(c)}E^{(c')}\sum_{\substack{n\ne 0 \\ n\ne\ell}} \sum_{k,k'} \tilde{C}_{n,k,k}^{(c)}C_{n-\ell,k'-\ell, k'-\ell}^{(c'), *} \nonumber \\
& +\overline{E}\sum_{c\ne 0}E^{(c)}\overline{E}^{(c)}\sum_{\substack{n\ne 0 \\ n\ne\ell}} \sum_{k, k'} D_{n,k,k'}^{(c)}D_{n-\ell,k-\ell, k'-\ell}^{(c), *}.
\label{eq:autocorr_v}
\end{align}
The crosscorrelation and pseudocorrelations of the $\{V_{m}\}$ are:
\begin{equation}
\left\langle V_{m}\overline{V}_{m+\ell}^*\right\rangle=\left\langle V_{m}V_{m+\ell}\right\rangle\color{black}=\left\langle V_{m}\overline{V}_{m+\ell}\right\rangle=0.
\label{eq:v_crosscorr}
\end{equation}

\subsubsection{Crosscorrelations}
For all $m,\ell$, we have
\begin{align}
& \left\langle \Psi_{m}\Theta_{\ell}\right\rangle=\left\langle \Psi_{m}\overline{\Theta}_{\ell}\right\rangle=0 \label{eq:theta_psi_uncorrelated}\\
& \left\langle V_{m}\Theta_{\ell}\right\rangle=\left\langle V_{m}\overline{\Theta}_{\ell}\right\rangle=0 \\
& \left\langle \Psi_{m}V_{\ell}^*\right\rangle=\left\langle \Psi_{m}V_{\ell}\right\rangle=\left\langle \Psi_{m}\overline{V}_{\ell}^*\right\rangle=\left\langle \Psi_{m}\overline{V}_{\ell}\right\rangle=0 \\
& \left\langle V_{m}X_{m}^*\right\rangle=\left\langle V_{m}X_{\ell}\right\rangle=\left\langle \overline{V}_{m}X_{m}^*\right\rangle=\left\langle \overline{V}_{m}X_{\ell}\right\rangle=0 \\
& \left\langle \overline{V}_{m}X_{\ell}^*\right\rangle=0
\end{align}
but for $m\ne \ell$ we also have
\begin{align}
\left\langle V_{m}X_{\ell}^*\right\rangle & =jE\sum_{c\ne 0}E^{(c)}\sum_{k}C_{\ell-m, k, k}^{(c)} \nonumber \\
& \quad +jE\sum_{c\ne 0}\overline{E}^{(c)}\sum_{k}\tilde{C}_{\ell-m, k, k}^{(c)}
\label{eq:isi}
\end{align}
so the additive NLI noise $V_{m}$ is correlated with the input symbols $X_{m}$ of the same polarization. This can be interpreted as inter-symbol interference (ISI).

\subsection{Large Accumulated Dispersion}
The synchronized case has $C_{n,k,k'}/2=\tilde{C}_{n,k,k'}=D_{n,k,k'}$. For IDA, $E^{(c)}=\overline{E}^{(c)}$ and $Q^{(c)}=\overline{Q}^{(c)}$ for all $c$, the approximations proposed in~\cite{dar2013properties} for large accumulated dispersion can be adapted to the 2-pol case. In other words, if $|\beta_2\Omega^{(c)}|\mathcal{L}/T\gg 1$ for all $c\in\mathcal{C}\setminus\{0\}$ then
\begin{align}
& \left\langle\Theta_{m}\right\rangle\approx 3\gamma\frac{\mathcal{L}}{T}\sum_{c\ne 0}E^{(c)} \\
& r_{\Theta}[\ell]\approx \frac{5\gamma^2\mathcal{L}}{T}\sum_{c\ne 0}\frac{Q^{(c)}-\left(E^{(c)}\right)^2}{\left|\beta_2\Omega^{(c)}\right|}\left[1-\frac{|\ell|T}{\left|\beta_2\Omega^{(c)}\right|\mathcal{L}}\right]^+
\label{eq:r_theta_analytic}
\end{align}
where $[x]^+\defeq\max(x, 0)$. The factors 3 and 5 are different than in \cite[Eqs.~(49) and (50)]{garcia2020mismatched} because the mean phase noise is $3/2$ times larger than its 1-pol counterpart due to the last summand in~\eqref{eq:E_theta}, and $r_{\Theta}[\ell]$ is $5/4$ times its 1-pol counterpart due to the last 2 lines in~\eqref{eq:r_theta}. Eqs.~\eqref{eq:crosscorr_theta_frac} and~\eqref{eq:r_psi_frac} still apply.

\section{Simplified models for computation}\label{sec:simplified}
In the following, the receiver removes the means $\left\langle\Theta_{m}\right\rangle$ and $\left\langle\overline{\Theta}_{m}\right\rangle$ of the phase noise, and we abuse notation and write $\Theta_{m}$ and $\overline{\Theta}_{m}$ for the resulting zero-mean variables. For each channel $c$, both polarizations have the same delay ($\Delta T^{(c)}=\overline{\Delta T}^{(c)}$), energy ($E^{(c)}=\overline{E}^{(c)}$), and fourth moment ($Q^{(c)}=\overline{Q}^{(c)}$). The models can be adapted to cases where these conditions do not hold but the resulting increase in the number of parameters substantially increases the computational cost of the training in Section~\ref{sec:estimation}.

\subsection{Polarization Drift (PD) Model}
The PD model was proposed in~\cite{czegledi2016polarization} and used in~\cite{secondini2019nonlinearity} to compute achievable rates of 2-pol systems. Consider the Pauli matrices:
\begin{equation}
\boldsymbol{\sigma}_1=\left(\begin{matrix}
0 & 1 \\ 1 & 0
\end{matrix}\right),\quad \boldsymbol{\sigma}_2=\left(\begin{matrix}
0 & -j \\ j & 0
\end{matrix}\right),\quad \boldsymbol{\sigma}_3=\left(\begin{matrix}
1 & 0 \\ 0 & -1
\end{matrix}\right).
\end{equation}
An \emph{isotropic random rotation on the Poincar\'e sphere} (IRRPS) of $\mathbf{v}_m\in\mathbb{C}^2$ can be represented as
$\mathbf{z}_m=\mathbf{U}_m\mathbf{v}_m$ where
\begin{align}
 & \mathbf{U}_m =\exp\left[j\left(\alpha_{1,m}\boldsymbol{\sigma}_1+\alpha_{2,m}\boldsymbol{\sigma}_2+\alpha_{3,m}\boldsymbol{\sigma}_3\right)\right] 
\label{eq:irr}
\end{align}
and where $\alpha_{1,m}$, $\alpha_{2,m}$ and $\alpha_{3,m}$ are i.i.d. real Gaussian with zero mean and variance $\sigma_A^2$. 
An IRRPS is such that the probability density function (PDF) of $\mathbf{z}_m$ does not change if $\mathbf{z}_m$ is expressed as a Stokes vector and rotated around $\mathbf{v}_m$.

The PD model for~\eqref{eq:y_1} is an extension of the Wiener phase noise model for the 1-pol case:
\begin{equation}
\mathbf{y}_m=e^{j\theta_m}\mathbf{J}_m\mathbf{x}_m+\mathbf{w}_m
\end{equation}
where $\{\Theta_m\}$ is Wiener phase noise that is common to both polarizations:
\begin{equation}
\theta_m=\theta_{m-1}+\sigma_\Delta\delta_m
\end{equation}
with the $\Delta_m$ being i.i.d. standard real Gaussian variables. The $\{\Theta_m\}$ do not change the polarization state of the signal. The matrix $\mathbf{J}_m$ is a random walk over the Poincar\'e sphere obtained as a cumulative product of IRRPSs of the form~\eqref{eq:irr}:
\begin{equation}
\mathbf{J}_{m}=\mathbf{U}_m\mathbf{J}_{m-1}.
\end{equation}
The NLI in the PD model has phase noise that converges to a uniform distribution over the Poincar\'e sphere~\cite{czegledi2016polarization}. This is not what the RP or LP models predict: the matrix $\mathbf{M}_m$ in~\eqref{eq:cpan_model} has entries that give a distribution with small variance around the transmitted symbol $\mathbf{x}_m$. Nonetheless, the PD model gives achievable rates that are very close to those obtained here, see~\cite{secondini2019nonlinearity} and the simulation results below.

\subsection{Markov Rotation (MR) Model}\label{sec:MR}
The statistical analysis in Section~\ref{sec:statistics} predicts in~\eqref{eq:r_thetas_synchr} that $\left\langle\Theta_{m}\Theta_{\ell}\right\rangle=(5/4)\left\langle\Theta_{m}\overline{\Theta}_{\ell}\right\rangle$. This can be modeled by choosing $\theta_{m}=2\phi_{m}+\overline{\phi}_{m}$ and $\overline{\theta}_{m}=2\overline{\phi}_{m}+\phi_{m}$, where the $\{\Phi_m\}$ and the $\{\overline{\Phi}_m\}$ are two independent zero-mean processes with variance $\langle\Theta_m^2\rangle/5$. We model the NLI rotation $\mathbf{M}_m$ of~\eqref{eq:M} as 
\begin{equation}
    \mathbf{M}_m=\exp\left[j\left(\begin{matrix}2\phi_m+\overline{\phi}_m & \psi_m \\ \psi_m^* & \phi_m+2\overline{\phi}_m
    \end{matrix}\right)\right]
    \label{eq:M_mr}
\end{equation}
where the $\Phi_{m}$ and $\overline{\Phi}_{m}$ are independent real Markov processes with memory $\mu$ and with the same statistical properties. Similar to what we did in~\cite{garcia2020mismatched}, for each $m$ we model $(\Phi_{m-\mu},\ldots,\Phi_{m})$ as jointly Gaussian with a symmetric Toeplitz covariance matrix $\mathbf{C}_\mu$ whose first column is $\left(r_{\Theta}[0],\ldots,r_{\Theta}[\mu]\right)^T/5$. This Markov approach yields long autocorrelation functions with small memory that replace the functions~\eqref{eq:r_theta}-\eqref{eq:r_bar_theta} from the analytical model.
We divide $\mathbf{C}_\mu$ into four blocks:
\begin{equation}
\mathbf{C}_{\mu}=\left(\begin{matrix}
c_{11} & \mathbf{c}_{12}^T \\ \mathbf{c}_{21} & \mathbf{C}_{22}
\end{matrix}\right)
\end{equation}
where $c_{11}$ is a scalar and $\mathbf{C}_{22}$ has size $\mu\times \mu$. As in~\cite{garcia2020mismatched}, the update function for $\phi_{m}$ is
\begin{equation}
\phi_{m}=\sum_{p=1}^{\mu}g_p\phi_{m-p}+\sigma_\mu\delta_{m}
\label{eq:phi_update}
\end{equation}
where the $\Delta_{m}$ are i.i.d. real standard Gaussian, where
\begin{equation}
\mathbf{g}^T =\left(\begin{matrix}
g_1 & \ldots & g_\mu
\end{matrix}\right)=\mathbf{c}_{12}^T \mathbf{C}_{22}^{-1}
\end{equation}
and where the increment variance $\sigma_\mu^2$ is
\begin{equation}
\sigma_v^2=c_{11}-\mathbf{g}^T\mathbf{c}_{21}.
\end{equation}

The $\{\Psi_m\}$ in~\eqref{eq:M_mr} is a proper complex Markov process independent of $\{\Phi_m\}$ and $\{\overline{\Phi}_m\}$, as predicted by ~\eqref{eq:theta_psi_uncorrelated}. The vector $(\Psi_{m-\mu},\ldots,\Psi_{m})$ has a conjugate symmetric Toeplitz covariance matrix whose first column is $\left(r_{\Psi}[0],\ldots,r_{\Psi}[\mu]\right)^T$ from~\eqref{eq:r_psi}. The process is generated similar to~\eqref{eq:phi_update}, but now the $\Delta_m$ are circularly symmetric complex Gaussian (CSCG) with unit variance.

\subsection{Simplified 2pCPAN Model}
As in~\cite{garcia2020mismatched}, we combine the ASE noise and the residual NLI noise~\eqref{eq:v} into one additive noise term $\mathbf{z}_{m}=\mathbf{w}_{m}+\mathbf{v}_{m}$. The simplified 2pCPAN model is
\begin{equation}
\mathbf{y}_{m}=\mathbf{M}_m\mathbf{x}_{m}+\mathbf{z}_{m}
\label{eq:dp_cpan}
\end{equation}
where $\mathbf{M}_m$ is given by~\eqref{eq:M_mr}. From~\eqref{eq:v_crosscorr}, the $Z_{m}$ are uncorrelated with the $\overline{Z}_{m}$. We model them as two CSCG processes independent of the $X_m$ and the $\overline{X}_m$ (i.e., we neglect~\eqref{eq:isi}) with autocorrelation functions 
\begin{equation}
r_{Z}[\ell]\defeq\left\langle Z_{m}Z_{m+\ell}^*\right\rangle=N_{\textrm{ASE}}\delta[\ell]+r_{V}[\ell]
\end{equation}
and $r_{\overline{Z}}$ defined similarly.

\section{Achievable Rates}\label{sec:rates}
The additive noise processes $\{Z_{m}\}$ and $\{\overline{Z}_{m}\}$ are uncorrelated with each other, and their autocorrelation functions seem to have a very small imaginary part. We thus whiten the noise processes separately for each polarization with real filters $\mathbf{h}$ and $\overline{\mathbf{h}}$ with $L$ taps and unit norm ($\|\mathbf{h}\|=\|\overline{\mathbf{h}}\|=1$):
\begin{align}
\mathbf{a}_m & =\sum_{\ell=0}^{L-1}\left(\begin{matrix}
h_{\ell} & 0 \\ 0 & \overline{h}_{\ell}
\end{matrix}\right)\mathbf{y}_{m-\ell} \nonumber \\
& = \left( \sum_{\ell=0}^{L-1}\left(\begin{matrix}
h_{\ell} & 0 \\ 0 & \overline{h}_{\ell}
\end{matrix}\right)\mathbf{M}_{m-\ell}\mathbf{x}_{m-\ell} \right) + \left(\begin{matrix} \xi_m \\ \overline{\xi}_m \end{matrix}\right).
\label{eq:mismatched_model}
\end{align}
Our mismatched channel model is~\eqref{eq:mismatched_model} with outputs $\{\mathbf{a}_{m}\}$, where $\mathbf{M}_m$ has the statistics given in Sec.~\ref{sec:MR}, and the $\{\Xi_{m}\}$ and $\{\overline{\Xi}_{m}\}$ are i.i.d. CSCG processes with variance $\sigma_{\Xi}^2$ estimated from training data.

As in~\cite{garcia2020mismatched}, we obtain a lower bound on the achievable rate of the Manakov channel via
\begin{equation}
I_q(\mathbf{X}; \mathbf{A})=h_q(\mathbf{A})-h_q(\mathbf{A}|\mathbf{X})
\label{eq:lb_theory}
\end{equation}
where $\mathbf{X}$ represents the two blocks of symbols $(X_1,\dots,X_M)$ and $(\overline{X}_1,\dots,\overline{X}_M)$, while $\mathbf{A}$ represents $(A_1,\dots,A_M)$ $(\overline{A}_1,\dots,\overline{A}_M)$. We define
\begin{equation}
h_q(\mathbf{A})=-\left\langle\log_2 q_{\mathbf{A}}(\mathbf{A})\right\rangle.
\label{eq:h_q}
\end{equation}
and similarly for $h_q(\mathbf{A}|\mathbf{X})$. Here $q_{\mathbf{A}|\mathbf{X}}(\cdot|\cdot)$ is the conditional distribution of the channel~\eqref{eq:mismatched_model}, and $q_{\mathbf{A}}(\mathbf{a})=\int_{\mathbf{X}} p_{\mathbf{X}}({\mathbf{x}})q_{\mathbf{A}|{\mathbf{X}}}(\mathbf{a}|{\mathbf{x}})\diff {\mathbf{x}}$ where $p_{\mathbf{X}}(\cdot)$ is the density of the input symbols $\mathbf{X}$. We choose the $\mathbf{X}$ to be i.i.d. CSCG  with variance $E=\mathcal{P}T$. The expectation in~\eqref{eq:h_q} is computed by Monte Carlo simulations and averaging over the channel realizations.

\subsection{Mismatched Output Entropy $h_q(\mathbf{A})$}\label{sec:h_Y}
Let $\underline{\mathbf{a}}=(a_{1},\ldots,a_{M})^T$ and $\overline{\underline{\mathbf{a}}}=(\overline{a}_{1},\ldots,\overline{a}_{M})^T$ be two blocks of simulated channel outputs. As $\mathbf{M}_m$ is unitary, the components of $\{\mathbf{M}_m\mathbf{X}_m\}$ are i.i.d. CSCG with variance $E$. This lets us separate the two polarizations. From~\eqref{eq:mismatched_model} we have
\begin{equation}
q_{{\underline{\mathbf{A}}}}({\underline{\mathbf{a}}})=\frac{\exp\left(-{\underline{\mathbf{a}}}^H\mathbf{R}_{\underline{\mathbf{A}}}^{-1}{\underline{\mathbf{a}}}\right)}{\det\left(\pi \mathbf{R}_{\underline{\mathbf{A}}}\right)}
\end{equation}
where $\mathbf{R}_{\underline{\mathbf{A}}}$ is a covariance matrix whose first column is $(r_{\underline{\mathbf{A}}}[0],\ldots,r_{\underline{\mathbf{A}}}[M-1])^T$ where
\begin{equation}
r_{\underline{\mathbf{A}}}[\ell] \defeq \left\langle A_{m}A_{m+\ell}^*\right\rangle=E\sum_{k=0}^{L-1}h_{k} h_{k+\ell}^* + \sigma_{\Xi}^2\delta[\ell].
\end{equation}
We set $h_{\ell}=0$ for $\ell\notin\left\{0,\ldots,L-1\right\}$. The mismatched output entropy is approximated as
\begin{equation}
h_q(\mathbf{A})=-\log_2 q_{\underline{\mathbf{A}}}(\underline{\mathbf{a}}) -\log_2 q_{\overline{\underline{\mathbf{A}}}}(\overline{\underline{\mathbf{a}}}).
\label{eq:h_U}
\end{equation}
We estimate $h_q(\mathbf{A})$ by averaging~\eqref{eq:h_U} over $N$ simulation runs.

\subsection{Mismatched Conditional Entropy $h_q(\mathbf{A}|\mathbf{X})$}
We use particle filtering~\cite{garcia2020mismatched} to estimate $h_q(\mathbf{A}|\mathbf{X})$. The method tracks the parameters $\phi_{m}$, $\overline{\phi}_m$ and $\psi_m$ using a list of $K$ particles. After processing $\mathbf{a}_{m-1}$, the $k$-th particle is a 4-tuple with a \emph{weight} $W_{m-1}^{(k)}$ and three realizations of the parameters of $\mathbf{M}_{m-1}$, namely a vector $(\phi_{m-\mu}^{(k)},\ldots,\phi_{m-1}^{(k)})$, a vector $(\overline{\phi}_{m-\mu}^{(k)},\ldots,\overline{\phi}_{m-1}^{(k)})$ and a vector $(\psi_{m-\mu}^{(k)},\ldots,\psi_{m-1}^{(k)})$. The $K$ weights sum to $1$. At the $m$-th iteration, the three realizations of each particle are updated using~\eqref{eq:phi_update} and the corresponding equation for $\Psi_m$. Let $p_{\boldsymbol{\Xi}}(\cdot)$ be the PDF of a two-component CSCG variable with variance $\sigma_{\Xi}^2$. The quantity
\begin{equation}
    D_m=\sum_{k=1}^{K}W_{m-1}^{(k)} p_{\boldsymbol{\Xi}}\left(\mathbf{a}_m-\sum_{\ell=0}^{L-1}\left(\begin{matrix}h_{\ell} & 0 \\ 0 & \overline{h}_{\ell}\end{matrix}\right) \mathbf{M}_{m-\ell}^{(k)}\mathbf{x}_{m-\ell}\right)
\end{equation}
gives an estimate of $q(\mathbf{a}_m | \mathbf{a}_1,\ldots,\mathbf{a}_{m-1},\mathbf{x}_1\ldots,\mathbf{x}_M)$. The weights are updated in a manner similar to~\cite[Eq. (70)]{garcia2020mismatched}:
\begin{equation}
W_m^{(k)}=\frac{W_{m-1}^{(k)}}{D_m}p_{\boldsymbol{\Xi}}\left(\mathbf{a}_m-\sum_{\ell=0}^{L-1}\left(\begin{matrix}h_{\ell} & 0 \\ 0 & \overline{h}_{\ell}\end{matrix}\right) \mathbf{M}_{m-\ell}^{(k)}\mathbf{x}_{m-\ell}\right).
\label{eq:weight_update}
\end{equation}
After the update~\eqref{eq:weight_update}, resampling~\cite{garcia2020mismatched} is applied if necessary. After the last iteration, the mismatched conditional entropy is estimated as
\begin{equation}
h_q(\mathbf{A}|\mathbf{X})=-\sum_{m=1}^{M} \log_2 D_m.
\label{eq:h_U_X}
\end{equation}
We refine the estimate of $h_q(\mathbf{A}|\mathbf{X})$ by averaging~\eqref{eq:h_U_X} over $N$ simulation runs. We then use~\eqref{eq:lb_theory} to lower bound the capacity.

\subsection{Estimating Model Parameters}\label{sec:estimation}
The parameters of the 2pCPAN model are computed from training data as in~\cite{garcia2020mismatched}. 
Similar to~\cite{secondini2019nonlinearity}, we estimate $\sigma_{\Xi}^2$ by neglecting the small correlations in $\{Z_m\}$ and $\{\overline{Z}_m\}$, and by approximating $\|\mathbf{Y}_{m}\|^2\big|\|\mathbf{X}_{m}\|^2$ as being independent with a noncentral chi-squared distribution with four degrees of freedom. More precisely, we estimate $\sigma_{\Xi}^2$ as
\begin{align}
& \hat{\sigma}_{\Xi}^2= \arg\max_{\sigma^2} \nonumber \\ & \sum_{m=1}^{M}\log\left[\frac{1}{\sigma^2}e^{-\frac{\|\mathbf{y}_{m}\|^2+\|\mathbf{x}_{m}\|^2}{\sigma^2}}\frac{\|\mathbf{y}_m\|}{\|\mathbf{x}_m\|}I_1\left(\frac{2\|\mathbf{y}_{m}\|\|\mathbf{x}_{m}\|}{\sigma^2}\right)\right].
\end{align}
where $I_1(\cdot)$ is the modified Bessel function of the first kind of order one. We estimate the mean phase noise $\langle\Theta_{m}\rangle$ as
\begin{equation}
\langle\hat{\Theta}_{m}\rangle=\textrm{angle}\left(\frac{1}{M}\sum_{m=1}^M y_{m}x_{m}^*\right).
\end{equation}
For the MR model with~\eqref{eq:M_mr}, we assume that $r_{\Phi}[\ell]=r_{\overline{\Phi}}[\ell]$ and that both autocovariance functions are proportional to~\eqref{eq:r_theta_analytic}. We also assume that $r_\Psi[\ell]$ is proportional to~\eqref{eq:r_theta_analytic}. We then minimize $h_q(\mathbf{Y}|\mathbf{X})$ (computed using particle filtering) over the two scaling factors. Finally, the whitening filters $\mathbf{h}$ and $\overline{\mathbf{h}}$ are chosen to be real, symmetric, equal, and with a length of $L=3$ taps per polarization. This leaves one real degree of freedom: $h_{2}=\overline{h}_2$. We minimize $h_q(\mathbf{A}|\mathbf{X})$ over $h_{2}$.

\section{Multi-Carrier Communication}\label{sec:mc}
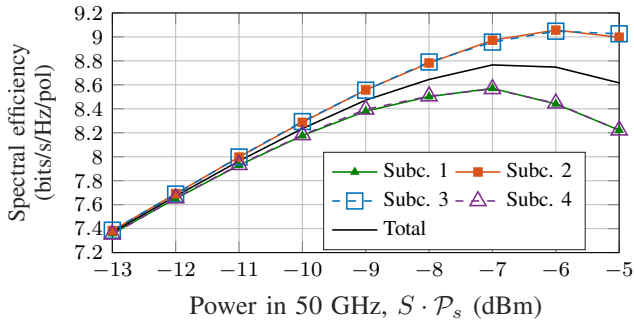
\begin{figure}[tbp]\centering
	\setlength{\figurewidth}{0.8\columnwidth}
	\setlength{\figureheight}{0.45\figurewidth}
%
%
\definecolor{mycolor1}{rgb}{0.00000,0.44700,0.74100}%
\definecolor{mycolor2}{rgb}{0.85000,0.32500,0.09800}%
\definecolor{mycolor3}{rgb}{0,0.5,0}%
\definecolor{mycolor4}{rgb}{0.49400,0.18400,0.55600}%
\begin{tikzpicture}

\begin{axis}[%
width=0.951\figurewidth,
height=\figureheight,
at={(0\figurewidth,0\figureheight)},
scale only axis,
xmin=-13,
xmax=-5,
xtick distance=1,
xlabel style={font=\color{white!15!black}},
xlabel={Power in 50 GHz, $S\cdot\mathcal{P}_s$ (dBm)},
ymin=7.2,
ymax=9.2,
ytick distance=0.2,
ylabel style={font={\small \color{white!15!black}}, align=center},
ylabel={Spectral efficiency\\ (bits/s/Hz/pol)},
grid,
axis background/.style={fill=white},
legend columns=2,
legend style={at={(0.675, 0.01)}, anchor=south, legend cell align=left, align=left, draw=white!15!black}
]
\addplot [semithick, color=mycolor3, mark=triangle*, mark options={scale=0.75}]
  table[row sep=crcr]{%
-13	7.35821311873072\\
-12	7.65066697264293\\
-11	7.92872380305678\\
-10	8.17650875416322\\
-9	8.38061851455032\\
-8	8.50169492312391\\
-7	8.57057469246276\\
-6	8.44087702747453\\
-5	8.22218709839757\\
-4	7.84135986621603\\
};
\addlegendentry{Subc. 1}

\addplot [semithick, color=mycolor2, mark=square*, mark options={scale=0.75}]
  table[row sep=crcr]{%
-13	7.3837093419516\\
-12	7.6918438744375\\
-11	7.99741106813902\\
-10	8.28742206118628\\
-9	8.55751779791184\\
-8	8.78243562920346\\
-7	8.97231711647212\\
-6	9.05690095913907\\
-5	8.99778009959569\\
-4	8.85426236878139\\
};
\addlegendentry{Subc. 2}

\addplot [semithick, dashed, color=mycolor1, mark=square, mark options={solid, scale=1.5}]
  table[row sep=crcr]{%
-13	7.38679941948804\\
-12	7.68870490261837\\
-11	7.9979962891103\\
-10	8.29364884880185\\
-9	8.55543009294444\\
-8	8.79087246904854\\
-7	8.95682988521217\\
-6	9.04799309436243\\
-5	9.02612404938834\\
-4	8.84375309611524\\
};
\addlegendentry{Subc. 3}

\addplot [semithick, dashed, color=mycolor4, mark=triangle, mark options={solid, scale=1.75}]
  table[row sep=crcr]{%
-13	7.35178567241307\\
-12	7.65421111084401\\
-11	7.93256216525676\\
-10	8.18164800349485\\
-9	8.39544171259054\\
-8	8.50535754313733\\
-7	8.5658909038045\\
-6	8.4457644049276\\
-5	8.22082854903412\\
-4	7.85416463895863\\
};
\addlegendentry{Subc. 4}

\addplot [semithick, color=black]
  table[row sep=crcr]{%
-13	7.37012688814586\\
-12	7.6713567151357\\
-11	7.96417333139071\\
-10	8.23480691691155\\
-9	8.47225202949929\\
-8	8.64509014112831\\
-7	8.76640314948789\\
-6	8.74788387147591\\
-5	8.61672994910393\\
-4	8.34838499251783\\
};
\addlegendentry{Total}

\end{axis}
\end{tikzpicture}%
	\caption{Spectral efficiency per subcarrier (Subc.) for a 1000-km link with uniform power allocation, synchronized channels, and the parameters in Table~\ref{tab:parameters}.}
	\label{fig:rates_per_sc}
\end{figure}

The second-order statistics of $\{\Theta_{m}\}$, $\{\overline{\Theta}_{m}\}$, $\{\Psi_m\}$ depend on the difference in frequency to the closest interfering channel~\cite{secondini2019nonlinearity, garcia2020mismatched}, and this observation motivates multi-carrier signaling~\cite{secondini2019nonlinearity}. Suppose each channel $c$ has $S$ subcarriers with bandwidth $\mathcal{B}/S$ and the subcarriers are back-propagated jointly at the receiver. We again use FDPA and apply power $\mathcal{P}_s$ to subcarrier $s$ for $s\inset{1}{S}$, see~\cite{garcia2020mismatched}. We here use the same heuristic to choose the powers as in~\cite{garcia2020mismatched}:
\begin{equation}
\underset{(\mathcal{P}_1,\ldots,\mathcal{P}_S)}{\arg\max} \sum_{s=1 }^S \textrm{rate}_s\left(\mathcal{P}_s\right)\textrm{s.t. }\sum_{s=1}^{S} \mathcal{P}_s=\mathcal{P}
\label{eq:fdpa_opt}
\end{equation}
where $\textrm{rate}_s(\mathcal{P}_s)$ is the rate of subcarrier $s$ with uniform power allocation and subcarrier power $\mathcal{P}_s$. The functions $\textrm{rate}_s(\cdot)$ are approximated by linear interpolation between simulated operating points at intervals of $1$ dBm, as shown in Fig.~\ref{fig:rates_per_sc}. The power allocation from~\eqref{eq:fdpa_opt} gives curves $\textrm{rate}_s(\cdot)$ that are used to optimize the powers in an iterative fashion. The results in Fig.~\ref{fig:rates} use one iteration of~\eqref{eq:fdpa_opt}.

\section{Numerical Results}\label{sec:results}
We simulate single-carrier and four-subcarrier (4SC) systems over a 1000-km fiber with the parameters in Table~\ref{tab:parameters}. The input distribution is i.i.d. Gaussian. The receiver uses a band-pass filter of bandwidth 50 GHz to isolate the COI, and then applies DBP jointly to $S$ subcarriers, followed by $S$ matched filters, and $S$ particle filters.

\subsection{Dual Polarization System}
In the 2-pol single-carrier system, we transmit 24 training sequences to estimate the parameters of the PD and 2pCPAN models according to Sec.~\ref{sec:estimation}. $N=120$ testing sequences are then used to compute achievable rates as explained in Sec.~\ref{sec:rates}. The sequence length is $M=6825$ symbols per WDM channel. 
In the 4SC system, the sequence length per subcarrier is $M=2047$ symbols, and there are 20 training sequences and $N=100$ testing sequences.

Frequency-dependent delays, but with $\Delta T^{(c)}=\overline{\Delta} T^{(c)}$ for all $c$, increase the rates with respect to the fully synchronized case. We chose the single-carrier delays as
\begin{align}
    \boldsymbol{\Delta}\mathbit{T}& =\left(\Delta T^{(-2)},\Delta T^{(-1)},\Delta T^{(0)},\Delta T^{(1)},\Delta T^{(2)} \right) \nonumber \\
    & =\left(5,6,-6,6,2\right)\frac{T}{15}.
    \label{eq:delays}
\end{align}
and the 4-subcarrier (4SC) delays as
\begin{align}
    & \boldsymbol{\Delta}\mathbit{T}_{\textrm{4SC}}=\left(\, (-25,-14,2,27),(27,-21,28,27),(-1,18,\right. \nonumber \\
    & \quad\left.-22,-5),(24,17,27,9),(-28,20,26,10)\, \right)\frac{T_{\textrm{4SC}}}{60}
    \label{eq:delays_sc}
\end{align}
where $T_{\textrm{4SC}}=4T$.
These delays were chosen randomly and optimization could increase the rates further. Note that here $\Delta T^{(0)}\ne0$.

The spectral efficiencies are plotted in Fig.~\ref{fig:rates}. The peak rate of the 4SC system with FDPA (4FDPA) using 2pCPAN is 8.91 bits/s/Hz/pol at -6 dBm. The 2pCPAN 4FDPA system provides a gain of 0.17 bits/s/Hz/pol or 0.8 dB with respect to our implementation of the system in~\cite{secondini2019nonlinearity} (PD, 4SC, same delays) at its peak rate. Starting with the system in~\cite{secondini2019nonlinearity}, frequency-dependent delays add 0.05 bits/s/Hz/pol, and FDPA adds another 0.09 bits/s/Hz/pol. Replacing the PD receiver with the MR receiver (without the whitening filter, not shown in Fig.~\ref{fig:rates}) adds 0.015 bits/s/Hz/pol, and adding the whitening filter~\eqref{eq:mismatched_model} adds another 0.015 bits/s/Hz/pol, yielding the 2pCPAN, 4FDPA curve.

\begin{figure}[tbp]\centering
	\setlength{\figurewidth}{0.8\columnwidth}
	\setlength{\figureheight}{0.46\figurewidth}
%
%
\definecolor{mycolor1}{rgb}{0.50000,1.00000,0.00000}%
\definecolor{mycolor2}{rgb}{0.00000,1.00000,0.25000}%
\definecolor{mycolor3}{rgb}{0.50000,0.00000,1.00000}%
\definecolor{mycolor4}{rgb}{0,0.5,0}%
\definecolor{mycolor5}{rgb}{1,0.5,0}
\definecolor{mycolor6}{rgb}{0,0.5,1}
\begin{tikzpicture}

\begin{axis}[%
width=0.976\figurewidth,
height=\figureheight,
at={(0\figurewidth,0\figureheight)},
scale only axis,
xmin=-10,
xmax=-5,
xtick distance=1,
xlabel style={font=\color{white!15!black}},
xlabel={Input power per 50-GHz channel (dBm)},
ymin=7.8,
ymax=9,
ytick distance=0.2,
grid,
ylabel style={font={\small \color{white!15!black}}},
ylabel={Spectral efficiency (bits/s/Hz/pol)},
axis background/.style={fill=white},
axis x line*=bottom,
axis y line*=left,
legend columns=3,
legend style={font=\scriptsize, at={(0.5, 1.05)}, anchor=south, inner sep=0, legend cell align=left, align=left, draw=white!15!black}
]
%

\addplot [semithick, color=red, mark=triangle*, mark options={solid, fill=red, red}]
table[row sep=crcr]{%
-10	8.25416247785239\\
-9	8.50751836713211\\
-8	8.71531365012873\\
-7	8.85759503995618\\
-6	8.90966384108473\\
-5	8.75960754784292\\
-4	8.35270306391901\\
};
\addlegendentry{2pCPAN, 4FDPA}

\addplot [semithick, color=mycolor6, mark=*, mark options={solid, fill=mycolor6, mycolor6}]
table[row sep=crcr]{%
-10	8.25831830940684\\
-9	8.49435985440966\\
-8	8.68961967779\\
-7	8.80282258355859\\
-6	8.80865234451839\\
-5	8.68088171990352\\
-4	8.3926669885939\\
};
\addlegendentry{2pCPAN, 4SC}

\addplot [semithick, color=mycolor4, mark=square*, mark options={solid, fill=mycolor4, mycolor4}]
table[row sep=crcr]{%
-10	8.20549720112627\\
-9	8.39569921520743\\
-8	8.55092851768051\\
-7	8.51580100906974\\
-6	8.35527665961602\\
-5	8.03496485496223\\
-4	7.78260368689372\\
};
\addlegendentry{2pCPAN}

\addplot [semithick, color=mycolor5, mark=triangle, mark options={solid, mycolor5}]
table[row sep=crcr]{%
-10	8.23468698464861\\
-9	8.48600831887307\\
-8	8.69123042736808\\
-7	8.83449386355324\\
-6	8.87997854095638\\
-5	8.74265490845772\\
-4	8.33067360597391\\
};
\addlegendentry{PD, 4FDPA}

\addplot [semithick, color=mycolor3, mark=o, mark options={solid, mycolor3}]
table[row sep=crcr]{%
-10	8.23772245024217\\
-9	8.47243516594244\\
-8	8.66564260569194\\
-7	8.78183106568852\\
-6	8.79023057915107\\
-5	8.65859911919704\\
-4	8.37370960534058\\
};
\addlegendentry{PD, 4SC}

\addplot [semithick, color=blue, mark=square, mark options={solid, blue}]
  table[row sep=crcr]{%
-10	8.19423816622202\\
-9	8.3909739445186\\
-8	8.51182436850436\\
-7	8.51058455979276\\
-6	8.34921330229005\\
-5	8.03200977414368\\
-4	7.58516334923455\\
};
\addlegendentry{PD}

\addplot [semithick, color=black]
  table[row sep=crcr]{%
-10	8.40885959646594\\
-9	8.74017910952567\\
-8	9.0716778580849\\
-7	9.40331911732814\\
-6	9.73507366456607\\
-5	10.0669182550719\\
-4	10.3988344044816\\
};
\addlegendentry{$\text{log}_\text{2}\text{ (1+SNR)}$}

\addplot [semithick, dashed, color=black]
table[row sep=crcr]{%
-10	7.96149902287613\\
-9	7.99742369925926\\
-8	7.9059091932411\\
-7	7.68809097576515\\
-6	7.33842092120104\\
-5	6.83143840821731\\
-4	6.28539164786909\\
};
\addlegendentry{Memoryless}

\addplot [semithick, color=gray]
table[row sep=crcr]{%
	-13	7.35982370030803\\
	-12	7.65965634834515\\
	-11	7.94971373845775\\
	-10	8.21809300415888\\
	-9	8.45470256900815\\
	-8	8.63177177102654\\
	-7	8.74278556891899\\
	-6	8.73006148360377\\
	-5	8.58272558241337\\
	-4	8.2944909092995\\
};
\addlegendentry{PD, 4SC, synchr.~\cite{secondini2019nonlinearity}}

\draw [thick] (-8, 8.725) rectangle (-6.9, 8.823);
\draw [thick, -Stealth] (-7.25, 8.68) -- (-6.9, 8.2);
\end{axis}

\begin{axis}[%
width=0.375\figurewidth,
height=0.2625\figureheight,
at={(0.725\figurewidth,0.02\figureheight)},
anchor=south,
scale only axis,
xmin=-8,
xmax=-6.9,
xtick={-8, -7.5, -7},
xticklabel pos=right,
ymin=8.725,
ymax=8.823,
ytick={8.75, 8.8},
grid,
axis background/.style={fill=white},
]
\addplot [semithick, color=red, mark=triangle*, mark options={solid, fill=red, red}]
table[row sep=crcr]{%
-10	8.25416247785239\\
-9	8.50751836713211\\
-8	8.71531365012873\\
-7	8.85759503995618\\
-6	8.90966384108473\\
-5	8.75960754784292\\
-4	8.35270306391901\\
};

\addplot [semithick, color=mycolor6, mark=*, mark options={solid, fill=mycolor6, mycolor6}]
table[row sep=crcr]{%
-10	8.25831830940684\\
-9	8.49435985440966\\
-8	8.68961967779\\
-7	8.80282258355859\\
-6	8.80865234451839\\
-5	8.68088171990352\\
-4	8.3926669885939\\
};
\addplot [semithick, color=mycolor5, mark=triangle, mark options={solid, mycolor5}]
table[row sep=crcr]{%
-10	8.23468698464861\\
-9	8.48600831887307\\
-8	8.69123042736808\\
-7	8.83449386355324\\
-6	8.87997854095638\\
-5	8.74265490845772\\
-4	8.33067360597391\\
};

\addplot [semithick, color=mycolor3, mark=o, mark options={solid, mycolor3}]
table[row sep=crcr]{%
-10	8.23772245024217\\
-9	8.47243516594244\\
-8	8.66564260569194\\
-7	8.78183106568852\\
-6	8.79023057915107\\
-5	8.65859911919704\\
-4	8.37370960534058\\
};

\addplot [semithick, color=gray]
table[row sep=crcr]{%
	-13	7.35982370030803\\
	-12	7.65965634834515\\
	-11	7.94971373845775\\
	-10	8.21809300415888\\
	-9	8.45470256900815\\
	-8	8.63177177102654\\
	-7	8.74278556891899\\
	-6	8.73006148360377\\
	-5	8.58272558241337\\
	-4	8.2944909092995\\
};

\legend{};

\end{axis}

\end{tikzpicture}%
	\caption{Spectral efficiency for a 2-pol 1000-km link with the parameters in Table~\ref{tab:parameters} and the frequency-dependent delays given by~\eqref{eq:delays} and~\eqref{eq:delays_sc}. The curve ``PD,4SC,synchr.'' has all delays equal to 0 and uses the same scheme as the best curve from~\cite[Fig. 3(a)]{secondini2019nonlinearity}.}
	\label{fig:rates}
\end{figure}

\subsection{Single Polarization System}
Frequency-dependent delays increase rates for the 1-pol case too. Fig.~\ref{fig:rates_1pol} is the result of the simulations in~\cite{garcia2020mismatched} with the single-carrier delays \eqref{eq:delays} and the 6-subcarrier (6SC) delays
\begin{align}
    & \boldsymbol{\Delta}\mathbit{T}_{\textrm{6SC}} =\left(\,(-37,-20,4,41,41,-31),(42,41,-2,27,\right. \nonumber \\ 
    & \quad\left.-33,-8),(37,26,41, 14,-42,31),(39,16,23,21\right. \nonumber \\ & \quad\left.-10,13),(-30,18,-43,-21,-41,-37)\,\right)\frac{T_{\textrm{6SC}}}{90}
    \label{eq:delays_6sc}
\end{align}
where $T_{\textrm{6SC}}=6T$. These delays were chosen randomly. The gain of our system (CPAN model, 6 SC, FDPA, and the delays in~\eqref{eq:delays} and \eqref{eq:delays_6sc}) over the best curve from~\cite{secondini2017fiber} (Wiener model, 6SC, no delays) is 0.19 bits/s/Hz in spectral efficiency or 1.15 dB in power efficiency.

\begin{figure}
    \centering
	\setlength{\figurewidth}{0.8\columnwidth}
	\setlength{\figureheight}{0.46\figurewidth}
%
%
%
\definecolor{mycolor1}{rgb}{0.50000,1.00000,0.00000}%
\definecolor{mycolor2}{rgb}{0.00000,1.00000,0.25000}%
\definecolor{mycolor3}{rgb}{0.50000,0.00000,1.00000}%
\definecolor{mycolor4}{rgb}{0,0.5,0}%
\definecolor{mycolor5}{rgb}{1,0.5,0}
\definecolor{mycolor6}{rgb}{0,0.5,1}
\begin{tikzpicture}

\begin{axis}[%
width=0.976\figurewidth,
height=\figureheight,
at={(0\figurewidth,0\figureheight)},
scale only axis,
xmin=-10,
xmax=-4,
xlabel style={font=\color{white!15!black}},
xlabel={Input power per 50-GHz channel (dBm)},
ymin=8,
ymax=9.2,
ytick distance=0.2,
grid,
ylabel style={font=\color{white!15!black}},
ylabel={Spectral efficiency (bits/s/Hz)},
axis background/.style={fill=white},
axis x line*=bottom,
axis y line*=left,
legend columns=3,
legend style={font=\scriptsize, at={(0.5, 1.05)}, anchor=south, inner sep=0, legend cell align=left, align=left, draw=white!15!black}
]

\addplot [semithick, color=red, mark=triangle*, mark options={solid, fill=red, red}]
  table[row sep=crcr]{%
-10	8.29541172887274\\
-9	8.57113840544248\\
-8	8.81934144755185\\
-7	9.02348780575743\\
-6	9.15459130469571\\
-5	9.17964282215977\\
-4	9.09653899832839\\
-3	8.45545127843484\\
};
\addlegendentry{CPAN, 6FDPA}

\addplot [semithick, color=mycolor6, mark=*, mark options={solid, fill=mycolor6, mycolor6}]
  table[row sep=crcr]{%
-10	8.29748462494311\\
-9	8.56360572145412\\
-8	8.80554804750769\\
-7	8.99023292535142\\
-6	9.09014957487479\\
-5	9.09587988877033\\
-4	8.96233305443135\\
-3	8.62701193232392\\
};
\addlegendentry{CPAN, 6SC}

\addplot [semithick, color=mycolor4, mark=square*, mark options={solid, fill=mycolor4, mycolor4}]
  table[row sep=crcr]{%
-10	8.25257841921529\\
-9	8.50325716393622\\
-8	8.68326432756765\\
-7	8.7932067130822\\
-6	8.77460733767028\\
-5	8.59541360974885\\
-4	8.27716517356223\\
-3	7.81759959104633\\
};
\addlegendentry{CPAN}

\addplot [semithick, color=mycolor5, mark=triangle, mark options={solid, mycolor5}]
  table[row sep=crcr]{%
-10	8.28748558192071\\
-9	8.56257267388884\\
-8	8.80948660272007\\
-7	9.01212281249527\\
-6	9.13753738177882\\
-5	9.15555025251243\\
-4	9.05738624260991\\
-3	8.38812280002158\\
};
\addlegendentry{Wiener, 6FDPA}

\addplot [semithick, color=mycolor3, mark=o, mark options={solid, mycolor3}]
  table[row sep=crcr]{%
-10	8.28957495442563\\
-9	8.55422988827904\\
-8	8.79480360728014\\
-7	8.9758645170107\\
-6	9.07067989842655\\
-5	9.06546855158443\\
-4	8.91755627913194\\
-3	8.53095070723769\\
};
\addlegendentry{Wiener, 6SC}

\addplot [semithick, color=blue, mark=square, mark options={solid, blue}]
  table[row sep=crcr]{%
-10	8.24602659030748\\
-9	8.49153403373197\\
-8	8.65695142685079\\
-7	8.74657398220103\\
-6	8.68519511885919\\
-5	8.4484874215286\\
-4	8.08705536560224\\
-3	7.57976327046832\\
};
\addlegendentry{Wiener}

\addplot [semithick, color=black]
  table[row sep=crcr]{%
-10	8.40885959646594\\
-9	8.74017910952567\\
-8	9.0716778580849\\
-7	9.40331911732814\\
-6	9.73507366456607\\
-5	10.0669182550719\\
-4	10.3988344044816\\
-3	10.7308074172433\\
};
\addlegendentry{$\text{log}_\text{2}\text{ (1+SNR)}$}

\addplot [semithick, dashed, color=black]
  table[row sep=crcr]{%
-10	8.07278280243399\\
-9	8.21899150936497\\
-8	8.21034021081668\\
-7	8.14510702899157\\
-6	7.90281373585884\\
-5	7.5278756522605\\
-4	7.0761930043548\\
-3	6.50789334431007\\
};
\addlegendentry{Memoryless}



\addplot [semithick, color=gray]
table[row sep=crcr]{%
	-13	7.3617299\\
    -12	7.6742852\\
    -11	7.9658629\\
    -10	8.2512127\\
    -9	8.5052656\\
    -8	8.7296929\\
    -7	8.8959932\\
    -6	8.9864593\\
    -5	8.9718213\\
    -4	8.8142074\\
};
\addlegendentry{Wiener, 6SC, synchr.~\cite{secondini2017fiber}}

\end{axis}

\end{tikzpicture}%
    \caption{Spectral efficiency for a 1-pol 1000-km link with the parameters in Table~\ref{tab:parameters} and the frequency-dependent delays given by~\eqref{eq:delays} and~\eqref{eq:delays_6sc}. The curve ``Wiener, 6SC, synchr.'' has all delays equal to 0 and uses the same scheme as the best curve from~\cite[Fig. 4(a)]{secondini2017fiber}.}
    \label{fig:rates_1pol}
\end{figure}

\section{Conclusions}
\label{sec:conclusions}
We extended the RP analysis of~\cite{mecozzi2012nonlinear} to two polarizations and derived a 2pCPAN model for WDM transmission and the Manakov equation. The model includes phase noise, polarization rotation, and additive noise. We applied particle filtering to derive lower bounds on the capacity of the 2-pol optical channel. With multiple carriers, FDPA, and frequency-dependent delays, we improved existing bounds by 0.17 bits/s/Hz/pol in spectral efficiency or 0.8 dB in power efficiency.

Further improvements might be possible by modeling~\eqref{eq:isi} as ISI, or by exploiting correlations of the phase noise and polarization rotation across subcarriers. Other ideas for future work are extensions to multi-mode communication and developing receivers that can achieve the reported gains.

\bibliographystyle{IEEEtran}
\bibliography{dpol_cpan.bib}

\end{document}